# Kinesin-12 KLP-18 contributes to the kinetochore-microtubule poleward flux during the metaphase of *C. elegans* one-cell embryo.

Nina Soler[1,*], Mathis Da Silva[1,*], Christophe Tascon[1], Laurent Chesneau[1], Pauline Foliard[1,†], Hélène Bouvrais[1], Sylvain Pastezeur[1], Loïc Le Marrec[2], Jacques Pecreaux[1,‡]

[1] CNRS, Univ Rennes, IGDR (Institut de Génétique et Développement de Rennes) – UMR 6290, F-35000 Rennes, France.

[2] CNRS, Univ Rennes, IRMAR (Institut de Recherche Mathématique de Rennes) – UMR 6625, F-35000 Rennes, France.

* Equally contributing authors

[†] Present address: IBGC, Bordeaux, France

[‡] Corresponding author: jacques.pecreaux@univ-rennes.fr

**Abstract:**

The mitotic spindle, a key structure to partition chromosomes during cell division, connects its poles to the chromosomes through microtubules. Their plus-ends, oriented towards the chromosomes, exhibit dynamic instability crucial for kinetochores' correct attachment. Involved in this process, the poleward flux implicates the displacement of microtubules towards the spindle poles, coordinated with polymerisation at the plus ends. The mechanisms behind this are diverse. It includes treadmilling powered by microtubule depolymerisation at the spindle poles, sliding of spindle microtubules by molecular motors like Kinesin-5, and pushing microtubules away from the chromosomes by chromokinesins. Interestingly, no such flux was reported in the *Caenorhabditis elegans* zygote, although all proteins contributing to flux in mammals have homologous in the nematode.

To explore this, we fluorescently labelled microtubules and conducted photobleaching. We found no global poleward flow; the bleached zone's edges moved inward. The centrosome-side front motion was caused by dynamic instability, while the chromosome-side front exhibited faster recovery, suggesting an additional mechanism. This extra velocity was localised near chromosomes, indicating that only kinetochore microtubules may undergo flux. Consistently, this flux depended on proteins ensuring the chromosome attachment and growth of the kinetochore microtubules, notably NDC-80, CLS-2$^{\text{CLASP}}$, and ZYG-9$^{\text{XMAP215}}$. Furthermore, this flux decreased as metaphase progressed and attachments transitioned from side- to end-on; it was reduced by SKA-1 recruitment. Traditional treadmilling was unlikely as many kinetochore microtubules do not reach spindle poles in the zygote spindle. Conversely, the depletion of kinesin-12 KLP-18$^{\text{KIF15}}$, which cross-links and focuses microtubules at meiosis, reduced the front rate. Ultimately, we propose that the sole kinetochore microtubules slide along spindle microtubules likely powered by KLP-18, contrasting with solid displacement in other systems, aligning with observations in human cells of decreasing flux with increasing chromosome distance.

# INTRODUCTION

During cell division, the mitotic spindle ensures a faithful partitioning of the genetic material into two identical sets distributed to each daughter cell. This specialised structure is assembled from microtubules (MTs) and their associated proteins (MAPs), especially molecular motors and cross-linkers. The microtubules are polarised semi-flexible polymers which alternate growth and shrinkage, and their plus ends are more dynamic than their minus ends (Akhmanova and Steinmetz, 2015). In the nematode C. *elegans* zygote, there are two types of microtubules in the spindle: kinetochore microtubules (kMT) with plus-ends anchored at the chromosomes and minus-ends spread along the spindle; and spindle microtubules (sMT) with minus-ends at the centrosomes and plus-ends not contacting the kinetochore (Muller-Reichert et al., 2010; Redemann et al., 2017). We included the branched microtubules along the sMT lattice in this latter category. Microtubule dynamic instability is essential for the spindle functions, a feature largely used in cancer therapies (Steinmetz and Prota, 2018; Vicente and Wordeman, 2019; Wordeman and Vicente, 2021). Numerous MAPs ensure a precise regulation of microtubule dynamics along the various phases of mitosis and spindle zones (Lacroix et al., 2018; Roostalu et al., 2018; Srayko et al., 2005). These molecular motors and MAPs stochastically bind and unbind to the microtubules, making the spindle highly dynamic (Elting et al., 2018; Nazockdast and Redemann, 2020).

An important aspect of ensuring spindle function is the attachment of the chromosomes by the kMTs while preserving the dynamic instability of the microtubule plus-end (Bakhoum and Compton, 2012). Thus, a specialised structure assembles at the centromeric regions of the chromosomes: the kinetochore (Cheeseman, 2014; Musacchio and Desai, 2017). In particular, the outer kinetochore protein NDC-80 makes the bridge with the microtubules (Cheeseman et al., 2004; Suzuki et al., 2016; Ye et al., 2016). In the nematode, chromosomes are holocentric; thus, kinetochores are scattered all along the chromosomes (Maddox et al., 2004). Chromosome attachment happens at first laterally, along the lattice of the microtubules (side-on), before being converted into an attachment at their plus-end (end-on) (Cheerambathur et al., 2013; Magidson et al., 2011). This latter can withstand the tension from the microtubules on the kinetochore. This tension is essential for the spindle assembly checkpoint (SAC) to identify chromosome misattachment and also for the subsequent mechanisms to fix these mistakes (Kuhn and Dumont, 2019; McVey et al., 2021). When attachments are correct, they are secured by the SKA complex, and the kMT dynamicity is reduced (Cheerambathur et al., 2013; Cheerambathur et al., 2017; Hanisch et al., 2006; Schmidt et al., 2012). MAPs regulate the dynamics of the attached microtubule like the rescue factor CLS-2$^{CLASP}$ (Al-Bassam and Chang, 2011; Cheeseman et al., 2005; Matthews et al., 1998). Some ubiquitous regulators also contribute, such as the polymerisation enhancer ZYG-9$^{XMAP215}$ (Matthews et al., 1998). In the nematode, like in other organisms, only one-fifth of the kMTs can directly reach the spindle pole (Kiewisz et al., 2022; Petry, 2016; Redemann et al., 2017). The remaining kMTs are likely connected to sMTs since their minus ends are very close to the sMT lattices.

Capturing and attaching chromosomes are partly stochastic, making errors possible. These are usually intercepted by the SAC (Nicklas, 1997). Notably and similar to mammalian embryos, in the C. *elegans* zygote, the SAC is weak, and some of its components are dispensable (Gerhold et al., 2018; Oegema and Hyman, 2006; Pintard and Bowerman, 2019; Tarailo et al., 2007). The tension exerted by the microtubules on the kinetochore involves their polymerising (Bakhoum et al., 2009; Cimini et al., 2006; Ertych et al., 2014; Lampson and Grishchuk, 2017; Schwietert et al., 2022). Combined with minus-end dynamics, it leads to poleward flux, which is a displacement of the microtubule network away from chromosomes (Barisic et al., 2021; Hotani and Horio, 1988; Mitchison, 1989; Steblyanko et al., 2020; Waters et al., 1996). A poleward flux was measured in various organisms, from Xenopus to mammals (Brust-Mascher et al., 2004; Brust-Mascher et al., 2009; Maddox et al., 2002; Maddox et al., 2003; Maiato et al., 2005; Matos et al., 2009; Yang et al., 2008). It plays an important role in correcting chromosome attachment errors, putatively by renewing the microtubules in contact with the kinetochore (Ertych et al., 2014; Ganem et al., 2005; Matos et al., 2009; Pereira and Maiato, 2012). During anaphase A, it also contributes to a synchronous migration of the sister chromatids toward their facing spindle poles (Ganem et al., 2005; Matos et al., 2009; Mitchison and Salmon, 1992). The most obvious flux mechanism is treadmilling, where the microtubule grows on the kinetochore side in a coordinated fashion with shortening at the spindle pole–this latter part powers the mechanism (Gaetz and Kapoor, 2004; Ganem et al., 2005). Such a mechanism was also recreated *in vitro* following a minimal system approach with only four proteins: XMAP215$^{ZYG-9}$, the microtubule plus-end binding EB1$^{EBP-2}$, CLASP2$^{CLS-2}$, and the microtubule depolymerising kinesin MCAK$^{KLP-7}$ (Arpag et al., 2020). This latter kinesin is essential to depolymerise the microtubules on the centrosome side. It can be assisted by severing enzymes like katanin, spastin and fidgetin (Zhang et al., 2007). Furthermore, treadmilling is consistent with the gel-like behaviour of the spindle reported in Xenopus egg extracts (Dalton et al., 2022) and the classic flux measurements using photobleaching/photoconversion experiments (Barisic et al., 2021; Mitchison and Salmon, 1992). Their results suggest a mostly solid motion of the spindle microtubule network at the steady state. Beyond the treadmilling mechanism, alternatives are twofold. Firstly, specialised molecular motors like kinesin-5 KIF11$^{BMK-1}$ and kinesin-12 KIF15$^{KLP-18}$ can slide the overlapping antiparallel microtubules carrying the kMTs through cross-linking (Miyamoto et al., 2004; Steblyanko et al., 2020; Uteng et al., 2008). Notably, kinesin-12 seems more specialised in parallel microtubules (Drechsler and McAinsh, 2016). Variants were proposed by sliding bridging fibres along each other (Jagric et al., 2021). The motion of these fibres is transmitted to other microtubules, including the kMTs and the sMTs, by cross-linking agents or motors like HSET$^{KLP-15/16/17}$, NuMA$^{LIN-5}$ or PRC1$^{SPD-1}$ (Elting et al., 2014; Risteski et al., 2022; Steblyanko et al., 2020). Secondly, chromokinesin KIF4A$^{KLP-19}$ congresses the chromosome arms and generates a reaction force that pushes the kMTs away from the chromosomes (Steblyanko et al., 2020; Wandke et al., 2012). It may play a similar role in nematode, although it would also regulate microtubule dynamics (Powers et al., 2004; Zimyanin et al., 2023). These various mechanisms can also be superimposed to ensure a robust poleward movement of the kMTs.

Surprisingly, although homologous proteins are present in the nematode C. e*legans,* no flux was reported in anaphase, while a putative but weak anti-poleward flux was suggested at metaphase (Labbe et al., 2004; Redemann et al., 2017). Notably, microtubules are highly dynamic in the nematode due to tubulin specificities (Chaaban et al., 2018). In contrast, a flux was found in zygote meiosis (Lantzsch et al., 2021). These paradoxes led us to investigate the flux during mitotic metaphase using the established fluorescence recovery after photobleaching (FRAP) (Axelrod et al., 1976; Giakoumakis et al., 2017; Matsuda and Nagai, 2014; White and Stelzer, 1999). Whereas we found no net flux for the spindle as a whole, we measured a poleward displacement of the microtubules close to the chromosomes. It depends on kinetochore proteins like CLS-2$^{CLASP}$, NDC-80 and the kinetochore-attachment states. We finally suggest that the kMTs be slid along the fixed sMTs by KLP-18$^{KIF15}$.

# RESULTS

## Two distinct microtubule-dynamics mechanisms account for recovering fluorescence after photobleaching.

We investigated the dynamics of the microtubules within the metaphasic spindle of the *C. elegans* zygote, studying recovery after photobleaching a 2.6μm wide region in the spindle (Methods § Microscopy). We used the GFP::TBB-2$^{β\text{-tubulin}}$ labelling. We previously reported no phenotype for this strain (Bouvrais et al., 2021). By analysing the average intensity in the photobleached region over time, we excluded that the diffusion of tubulin dimers accounted for the recovery inside the spindle (Supplemental Text §1 and Fig S1AB). Next, to investigate a putative flux, we set out to analyse the shape of the bleached area over time (Movie S1, Fig 1A, B). We used the centrosome on the bleaching side as a spatial reference to register images and get rid of the effects of spindle pole motion (Methods § Image processing, Fig S2). We produced the kymograph of individual embryos along the spindle axis by median projection along the transverse axis (Fig S3). The fast dynamics of the microtubules imposed imaging at 12.5 frames per second, leading to a low signal-to-noise ratio (SNR). To account for low SNR, we averaged the kymographs from several embryos after aligning them on the centrosome-side top corner of the bleached region in the kymograph and matching their intensity histograms. We then segmented the bleached region of the averaged kymograph. We fitted linearly its boundaries to quantify the displacement of the two fronts over time by their slope in the kymograph (Fig 1D, blue lines). We used a leave-one-out resampling (Jackknife) to obtain the standard errors of the slopes (Efron and Tibshirani, 1993; Quenouille, 1956). Such an approach also enabled us to visualise the front-slope distributions by displaying N-1 averages (Fig 1C, Methods § Statistics on kymograph front slopes). We finally computed the slope of the mid-curve between the edges corresponding to the bleached region solid displacement (Fig 1D, black line). It read 0.017 ± 0.028 μm/s, oriented poleward ($N = 10$, mean ± standard error (se) obtained by Jackknife resampling, $p = 0.56$ compared to 0). We concluded that the bleached area underwent no global displacement, meaning no significant global flux. This is consistent with previous reports on the lack of flux in metaphase and anaphase (Labbe et al., 2004; Redemann et al., 2017).

However, a visual inspection of the kymograph (Fig 1B, D) indicated a fluorescence recovery by closing the edges through two inward-oriented fronts. We investigated this V-shaped recovery and measured the closure velocities (Fig 1C). We read 0.117 ± 0.028 μm/s on the chromosome side, which significantly differed from 0 ($N = 10$, $p = 0.002$) and 0.082 ± 0.039 on the centrosome side ($p = 0.05$), both oriented inwards. These values of the slope were significantly different from each other ($p = 1.9\times10^{-5}$, $N = 10$). We reckoned that these two fronts of opposite directions might reveal distinct microtubule dynamics close and far from the chromosomes. Indeed, we expect the kMTs to be numerous enough to be detected close to chromosomes but vastly overwhelmed by sMTs at the location of the centrosome front (Redemann et al., 2017).

To safeguard against artefacts, we repeated the experiment using photoconversion (see Methods § Microscopy). We crossed strains with mEOS3.2::TBB-2$^{β\text{-tubulin}}$, labelling microtubules, and mCherry::TBG-1$^{γ\text{-tubulin}}$, revealing the centrosomes. We used this second label to register the images over time, as reported above (Methods § Image processing). We produced the kymographs, and again, to cope with the low SNR, we averaged them across embryos after aligning them on the photoconverted region. We obtained a similar V-shaped pattern (Fig S1C) as expected. We concluded that the two fronts were not artefactual.

After validating that the motion of the two closing edges depended on microtubule dynamics using *zyg-9*$^{XMAP215}$*(RNAi)*, we investigated the front on the centrosome side (Suppl Text §2, Fig S1D, E, replica in Fig S1F). This depletion decreases the polymerisation rate at the plus ends. The most obvious possibility of accounting for the observed fronts is a flux of the microtubules themselves. While a poleward flux of kMTs was not observed in the nematode, it is well-known in numerous other species. In contrast, an anti-poleward flux of sMTs appeared much more unlikely. The edge of the bleached region on the centrosome side was located at about half the distance between the chromosomes and the spindle pole; at this place, the sMTs are three times more numerous than the kMTs (Redemann et al., 2017). We thus focused on the sMTs and reckoned that the front displacement on the centrosome side might be accounted for by the dynamics of the sMTs (growth and shrinkage) combined with diffraction due to imaging. We modelled this phenomenon and observed a closing front similar to the experimental one (Suppl Text §3, Fig S4, S5). The reduction in the centrosome-side front velocity as the spindle got longer supported the proposed mechanism. Notably, this latter phenomenon is expected to cause two symmetric closing edges, i.e., on both the centrosome and chromosome sides, that move at the same speed. We concluded that dynamic instability is sufficient to account for the front motion on the centrosome side. However, because no correlation was visible between the chromosome side velocity and the spindle length (Fig S6), and because recovery was faster on that side, it suggested an additional mechanism.

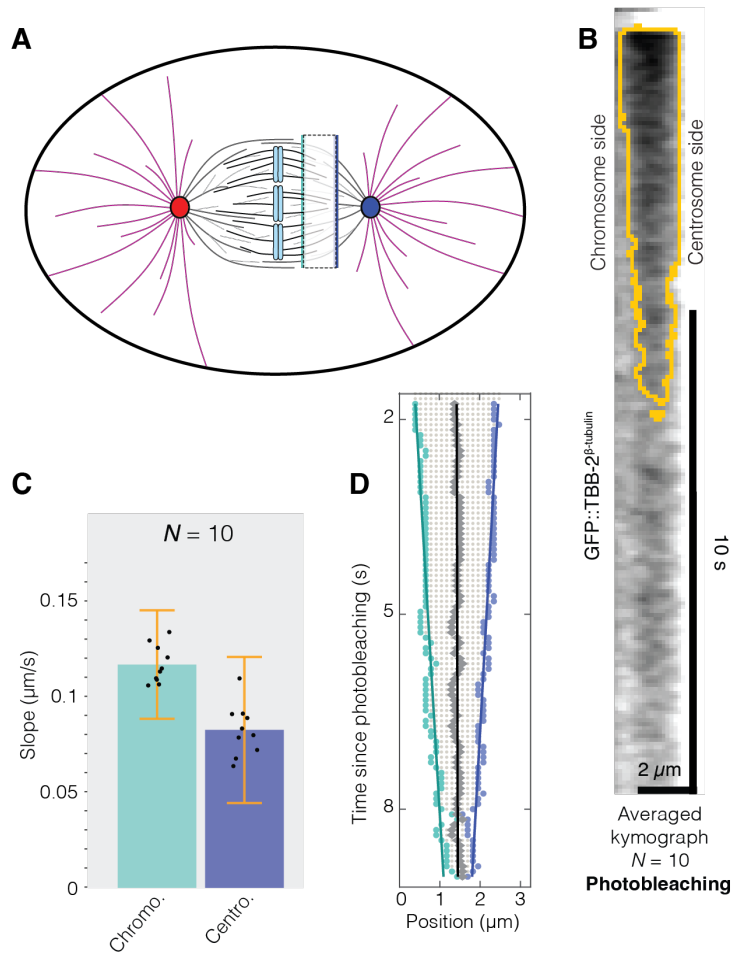

**Figure 1: Tubulin fluorescence recovery after photobleaching in the metaphasic spindle of *C. elegans*.** (**A**) Schematics of the FRAP experiment to measure front displacements. Thick dark grey lines depict the spindle microtubules emanating from (red) anterior and (blue) posterior centrosomes; thin light grey ones, the one branching from other microtubule lattices. Black lines correspond to kinetochore microtubules bound to (blue bars) the condensed sister chromatids. Astral microtubules are depicted in purple colour. The dashed box corresponds to the bleached area. The blue bars represent the measurement location of the fronts on (dark blue) the chromosome and (light blue) the kinetochore sides. (**B**) Averaged kymograph over the posterior half-spindle of $N = 10$ GFP::TBB-2$^{\beta\text{-tubulin}}$ labelled embryos. The centrosome was located on the right-hand side. The orange line delineates the bleached region as obtained by our analysis (Movie S1, Methods § Image processing). (**C**) Front velocities by analysing the segmented bleached-region of $N = 10$ GFP::TBB-2$^{\beta\text{-tubulin}}$ non-treated embryos. Black dots represent the average of $N$-1 embryos, leaving out each embryo in turn (Methods § Statistics on kymograph front slopes). Bars correspond to means; error bars are estimated standard errors using Jackknife resampling. Light blue bars are velocities on the chromosome side, and dark blue the ones on the centrosome side. It corresponds to the kymograph analysis displayed in B. (**D**) Linear fits of the edges of the segmented kymograph averaged over the non-treated embryos reported in panels B and C. Grey dots depict the pixels in the bleached region. Boundaries are highlighted by coloured dots on (turquoise, left edge) the chromosome side and (blue, right edge) the centrosome side. Grey diamonds depict the region mid-line. Linear fits are reported as lines of corresponding colours.

## A poleward flux of kinetochore microtubules may account for the chromosome-side recovery.

We reckoned that the higher front rate velocity observed on the chromosome side and the lack of correlation between the spindle length and the recovery front velocity suggested that a second mechanism might superimpose specifically on the chromosome side. The most obvious addition would involve the kMTs growing at their plus ends, i.e. at the chromosomes, while sliding towards the poles and thus along the spindle microtubules. Consistently, depletion of ZYG-9$^{XMAP215}$ reduced the front velocity on the chromosome side (Fig S1D, E). This hypothesis was consistent with Redemann and colleagues not measuring this poleward flux because being far from the chromosomes. Indeed, the kMTs are abundant compared to the sMTs, only close to the chromosomes (Redemann et al., 2017).

In other organisms, the poleward flux involves all the microtubules, although their dynamics may differ (Conway et al., 2022; Risteski et al., 2022; Zhai et al., 1995). Dalton and co-authors proposed that the spindle undergoes gelation to account for the solid poleward motion of the microtubules within spindles prepared from *Xenopus laevis* extracts (Dalton et al., 2022). In contrast, we proposed that only the kMTs would undergo a poleward flux. As their number decreases with distance from the kinetochore (Redemann et al., 2017), the flux could be seen as restricted close to the kinetochores. To challenge this idea, we performed FRAP experiments, bleaching a smaller area located either close or far from the spindle pole (Fig 2A). When looking far from the chromosomes, we obtained similar front velocities on edges facing chromosomes and centrosome (Fig 2B, maroon background). It suggested that, in this spindle region, both fronts reflected similar mechanisms. Moreover, this symmetry is incompatible with sMTs undergoing a flux in either direction. Because the sMTs are in the vast majority compared to the kMTs, the recovery was likely due solely to the microtubule dynamics combined with imaging diffraction, as detailed above. Close to the kinetochore, we measured higher front velocity on the chromosome side than the centrosome one, which is consistent with our previous measurement. However, this latter value was larger than in the bleached region far from the centrosomes. It may be related to a gradient of RAN protein around the chromosomes to promote microtubule growth (Bamba et al., 2002; Srayko et al., 2005). Overall, this experiment was consistent with our hypothesis that only the kMTs underwent poleward flux. It also accounted for the difference between our findings and previous measurements because the kMT flux could only be seen close to the chromosomes (Labbe et al., 2004; Redemann et al., 2017). Anywhere else, the non-fluxing sMTs, which are in the majority, hide the kMTs flux.

To further ascertain that fluxing kMTs could cause the front motion on the chromosomes side, we set out to genetically impair the kMTs' attachment to the kinetochore. We partially depleted NDC-80 in such a condition that chromosome segregation eventually succeeded in anaphase. We observed that the spindle in metaphase elongated during early metaphase (further termed early elongation) and plateaued before the anaphase onset, as previously reported (Fig S7A) (Cheerambathur et al., 2013). We studied the microtubule dynamics by FRAP during both phases (Fig 2C, D). During the early elongation, the front velocity on the chromosome side increased, while during the plateau, it decreased compared to the control. Such a result suggested that the kinetochore-microtubule attachment was involved in setting the corresponding front velocity. We repeated the experiment and obtained similar results (Fig S7C). Our partial depletion of NDC-80 led to a delay in microtubule attaching end-on to the kinetochores (Cheerambathur et al., 2013; Cheerambathur et al., 2017; Cheeseman et al., 2004; Lange et al., 2019). Thus, kinetochore-microtubule attachments could not bear a normal load, leading to early elongation. It might account for the observed increased front velocity under the hypothesis that the kMTs underwent a poleward flux. Later, during the plateau, we observed a velocity similar to or slightly reduced compared to the control. Altogether, these observations were suggestive that growing kMTs at the kinetochores may be necessary for the mechanism causing the chromosome recovery front.

In the classic poleward-flux mechanism, CLS-2$^{CLASP}$ is critical to ensure that the microtubules switch to polymerisation while undergoing flux (Arpag et al., 2020; Cheeseman et al., 2005). We thus partially depleted this protein. We observed a shorter spindle during metaphase, as expected (Cheeseman et al., 2005) (Fig S7B, F). We monitored the front velocity using our FRAP assay and found an increased velocity on the chromosome side that recalled the one observed during the precocious spindle elongation upon *ndc-80(RNAi)* (Fig S8, S7BD). We interpreted it similarly: the kMTs were pulled out from the kinetochores upon *cls-2(RNAi)*, causing the increased front velocity. Consistently, upon strong depletion, the spindle breaks due to the cortical pulling forces (Cheerambathur et al., 2017; Cheeseman et al., 2005). Overall, we concluded that our results, especially the observed recovery front on the chromosome side, could be explained under the hypothesis that the kinetochore-microtubule fluxing towards the poles, requiring microtubule polymerisation at the plus-ends.

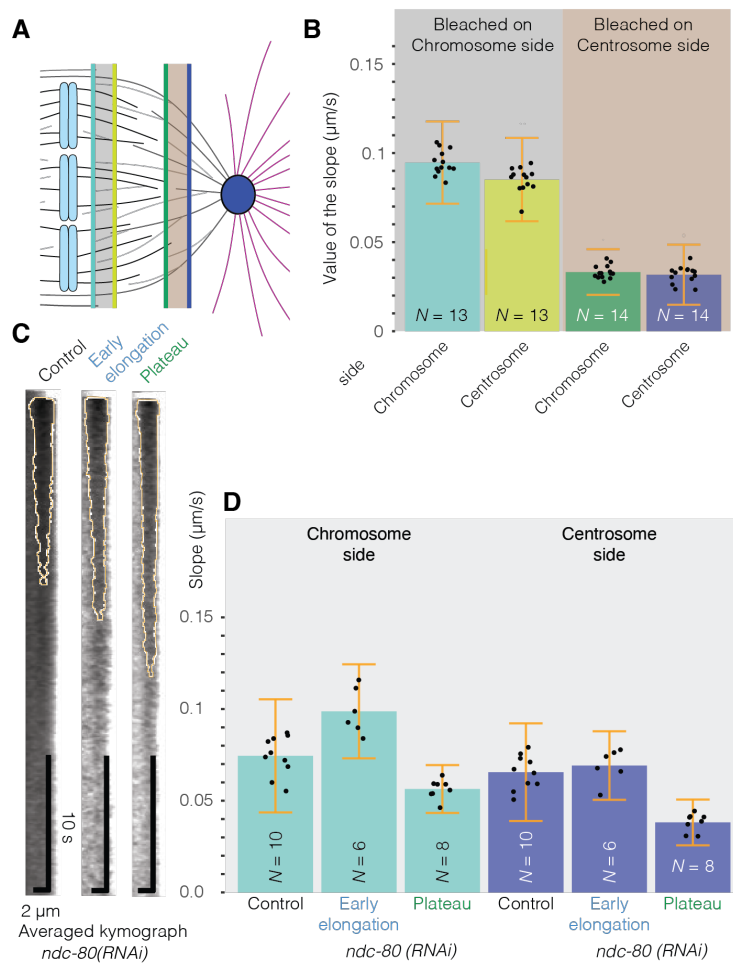

**Figure 2: Front velocity on the chromosome side requires kMT polymerising at the plus-ends.** (**A**) Schematics of the FRAP experiment, with bleaching localised close to either the chromosomes or the centrosome to measure front displacement during recovery. Grey lines depict the spindle microtubules emanating from (blue) the posterior centrosome. Black lines correspond to the kinetochore microtubules bound to (blue bars) the condensed sister chromatids. The grey and maroon boxes are the bleached areas close to chromosomes and the centrosome, respectively. The coloured lines depict the approximate positions of the front measurements corresponding to B. (**B**) Front velocities by segmenting the dark regions of non-treated embryos, bleached either (grey background, $N = 13$) close to chromosomes or (maroon background, $N = 14$) close to centrosomes. Microtubules were labelled using GFP::TBB-2$^{\beta\text{-tubulin}}$. The colour of the bars corresponds to the schematics (A). (**C**) Averaged kymographs over the posterior half-spindle for the data presented in (D), with centrosome on the right-hand side. The orange line delineates the bleached region as obtained by our analysis (Methods § Image processing). (**D**) Front velocities by segmenting the dark region of $N = 6$ *ndc-80(RNAi)* embryos bleached during early elongation, $N = 8$ *ndc-80(RNAi)* bleached during spindle length plateau and their $N = 10$ controls (typical micrograph at Fig S7A). Microtubules were labelled using GFP::TBB-2$^{\beta\text{-tubulin}}$. Light blue bars are values on the chromosome side and dark blue on the centrosome side. The experiment was replicated (Fig S7C). In panels B and D, black dots represent averages of $N$-1 embryos, leaving out, in turn, each embryo (Methods § Statistics on kymograph front slopes). Bars correspond to means, and error bars are estimated using Jackknife resampling.

### The chromosome-side poleward flux depends on the attachment status at the kinetochore.

We reckoned that during spindle assembly, the kinetochores first attached laterally, with reduced resistance to tension on kinetochores, then end-on (Cheerambathur and Desai, 2014). This second configuration couples the dynamics of the kMT plus-end with the tension (Cheeseman et al., 2004; Suzuki et al., 2015). Next, the SKA complex is recruited during late metaphase and reduces the dynamicity of the attachments (Cheerambathur et al., 2017; Schmidt et al., 2012). We thus grouped our non-treated and control experiments into 15s-long bins by the time the bleaching was performed and measured the front velocities. On the chromosome side, we found that it significantly decreased over time (Fig 3A). It suggested that the dynamics of the kMT plus-ends might impact the measured velocity of the chromosome-side front, which aligned with our hypothesis. We also excluded that the observed correlation could result from modest variations in the distance to the chromosomes (Fig S8B). Notably, the front of the centrosome side also displayed a decreasing velocity (Fig S8A). In the framework of the proposed model for this side (Suppl Text §3), we attributed that to either the decreasing microtubule growth rate over metaphase and anaphase, or the spindle getting slightly longer over the metaphase (Goshima and Scholey, 2010; Srayko et al., 2005). Interestingly, we observed decreased front velocities on both sides during early anaphase compared to metaphase (Fig S9A). It can be consistent with locking correct chromosome attachment at the kinetochores and the reduced microtubule growth rate at anaphase (Asbury, 2017; Srayko et al., 2005).

To exclude that a cell-scale change of microtubule dynamics over time accounted for the above results, we set out to manipulate the dynamics of the kinetochore microtubules genetically, although maintaining their attachments. We depleted the protein SKA-1 by RNAi, preventing the locking of the end-on attachments. Significantly and in contrast to mammalian cells, it did not delay anaphase onset (Schmidt et al., 2012). We measured the velocity of the fronts in the time range from 60 to 30 s before anaphase onset and observed a faster rate in depleted embryos(Fig 3B, replicated in Fig S9B). Indeed, the recruitment of this protein is expected at that moment(Cheerambathur et al., 2017). This was expected in our hypothesis of a chromosome-side front motion caused by kMT poleward flux.

We then performed the converse experiment, getting precocious and stronger recruitment of SKA complex, and thus a decrease of kinetochore-microtubule dynamics. To do so, we used the previously published mutated *ndc-80,* where four Aurora kinase phosphorylation sites were changed to alanines and termed NDC-80-4A (Cheerambathur et al., 2017). The authors made it RNAi-resistant by changing some triplets to synonymous codons. We also considered the strain, with similar scrambling of gene coding but no alteration of the phosphorylation sites as control, termed NDC-80-scr. It enabled us to target the endogenous copy by RNAi. We crossed these strains with the one carrying the GFP::TBB-2$^{\beta\text{-tubulin}}$ labelling and measured the flux in the time frame from 120 to 30 seconds before anaphase onset. We observed a decreased chromosome-side front velocity in NDC-80-4A conditions with respect to NDC-80-scr, both upon *ndc-80(RNAi)* or control RNAi. This was expected, considering that SKA decreased the kMT dynamics at the kinetochore too early or too strongly (Fig 3C). In contrast, on the centrosome side, upon depleting the endogenous NDC-80 protein and rescuing with the mutated protein NDC-80-4A, we observed an increase in front velocity, suggesting that the precocious recruitment of SKA affected differentially the front on each side (Fig S9C).

Altogether, the timely evolution of the velocity of the chromosome-side front and its dependence upon the SKA complex confirmed that the kinetochore microtubules undergoing a poleward flux likely contributed to the chromosome front motion.

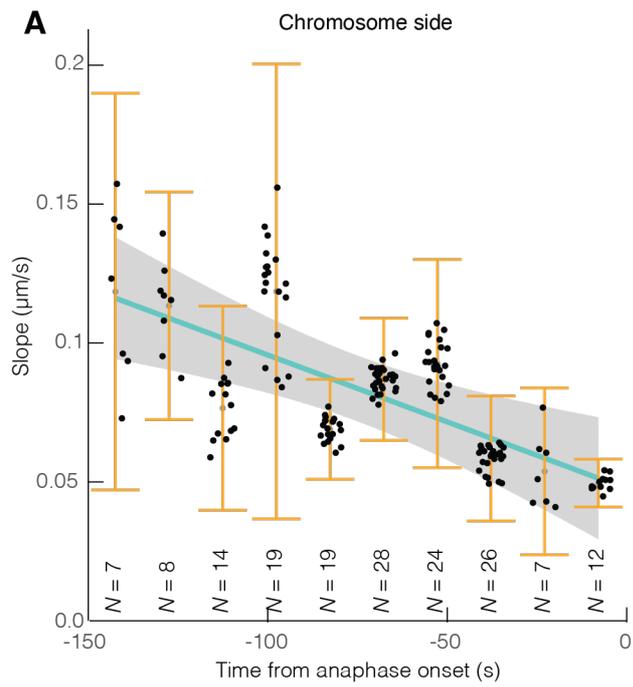

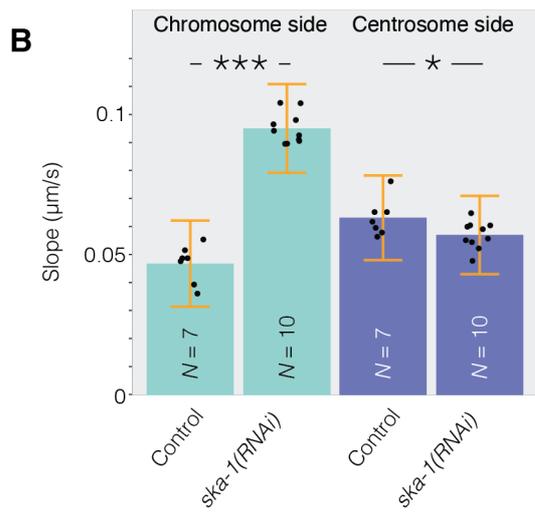

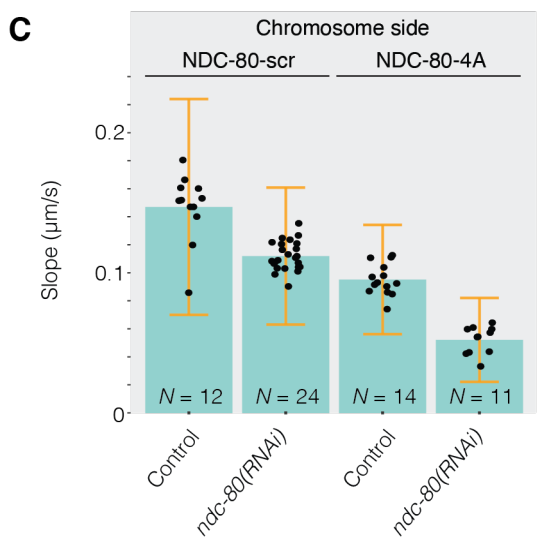

**Figure 3:** (A) Correlation between the time of bleaching and the front velocities by segmenting the dark regions of non-treated or control embryos of experiments reported in other figures. Microtubules were labelled using GFP::TBB-2$^{\beta\text{-tubulin}}$. Black dots represent averages of N-1 embryos, leaving out, in turn, each embryo (Methods § Statistics on kymograph front slopes). Error bars are estimated using Jackknife resampling. Pearson coefficient read $R$ = -0.82 ($p$ = 0.0037). The grey-shaded region corresponds to the confidence interval at 0.95 on the correlation-line coefficients. (B) Front velocities by segmenting the dark region of $N$ = 7 *ska-1(RNAi)* embryos bleached during late metaphase (-60 to -30 s from anaphase onset) and their $N$ = 10 controls. The experiment was replicated (Fig S9B). (C) Front velocities on chromosome side by segmenting the dark region of NDC-80-scr embryos either ($N$ = 12) treated by *ndc-80(RNAi)* or ($N$ = 24) control and bleached; NDC-80-4A embryos either ($N$ = 14) treated by *ndc-80(RNAi)* or ($N$ = 11) control. Corresponding velocities on the centrosome side are in Fig S9C. In panels (B and C), microtubules were labelled using GFP::TBB-2$^{\beta\text{-tubulin}}$. Black dots represent averages of N-1 embryos, leaving out, in turn, each embryo (Methods § Statistics on kymograph front slopes). Bars correspond to means; error bars are estimated standard errors using Jackknife resampling. Light blue bars are velocity values measured on the chromosome side and dark blue on the centrosome side.

## The tetrameric kinesin KLP-18 is likely sliding the kinetochore microtubules along the spindle microtubules.

Which mechanism could account for the kMT poleward flux? The most classic one is treadmilling, powered by the depolymerisation at the minus end. Electronic micrographs reported that about a quarter of the kMTs reached their corresponding pole (Redemann et al., 2017). Besides, half of the microtubules had opened minus-ends at the centrosomes (O'Toole et al., 2003). The treadmilling mechanism usually involves MCAK as the main microtubule depolymeriser (Arpag et al., 2020; Brust-Mascher et al., 2004). The MCAK homologous protein, KLP-7, is also localised at the kinetochores and spindle poles (Han et al., 2015; Sarov et al., 2012). We depleted it and measured the front velocity again (Fig S10). We did not observe a decreased front velocity on the chromosome side. The trend was instead an increase in velocity that we attributed to a role of KLP-7 at the kinetochore, putatively facilitating the turnover of the kMTs (Jaqaman et al., 2010; Kline-Smith et al., 2004; Wordeman et al., 2007). *In vitro*, a minimal treadmilling system comprises CLASP$^{CLS-2}$, XMAP215$^{ZYG-9}$, MCAK$^{KLP-7}$ and EB1$^{EBP-2}$. While the former two proteins were involved in causing the front velocity that we observed on the chromosome side, the depletion of the latter two did not result in significant phenotype (Suppl Text §5, Fig S10 and S11). In conclusion, we considered treadmilling an unlikely mechanism. We rather suggested that the kMTs moved locally poleward in a mechanism reminiscent of the poleward flux despite these microtubules being too short to reach the centrosome.

As an alternative, we considered the sliding of the kMTs along the sMTs, which were not moving. KLP-18, a kinesin-12 motor, appeared as a plausible candidate. Its mammal counterpart, KIF15, was shown to slide parallel microtubules *in vitro* (Drechsler and McAinsh, 2016). Furthermore, KLP-18 is also essential to correctly align the microtubules in the acentrosomal meiotic spindle of the nematode (Cavin-Meza et al., 2022; Wolff et al., 2022). Importantly, KLP-18 is localised in the mitotic spindle (Suppl Video S2). We thus set out to deplete this protein by RNAi. Because it is required for successful meiosis, conditions were hypomorphic. We measured a modest, although significant, decrease in the chromosome-side front velocity, while no effect was observed on the centrosome side (Fig 4A). It suggested that KLP-18 could contribute to kMTs poleward fluxing.

To ascertain the role of this kinesin and get a stronger phenotype, we considered the temperature-sensitive mutant *klp-18(or447ts)* compared to OD868 denoted *klp-18(+)* (Connolly et al., 2014). After culturing the worms at 15°C, they were either imaged at 15°C (permissive condition) or incubated at 25°C for 30 min prior to dissection and imaging (restrictive). In this latter condition, we selected the embryo with no apparent pronuclei defect. We measured the velocity of the front on the chromosome side and observed a decrease by a factor of two in embryos carrying the mutated protein compared to control ones (Fig 4B). It is noteworthy that the microtubule growth-rate increases with temperature, accounting for the observed increase in front velocity between permissive and restrictive temperatures (Chaaban et al., 2018). Notably, on the centrosome side at a given temperature, we observed no apparent difference between the control and mutated KLP-18 conditions (Fig S12). We concluded that KLP-18 specifically contributed to the front velocity on the chromosome side. We proposed that this tetrameric kinesin slid the kinetochore microtubules along the spindle ones, remaining immobile.

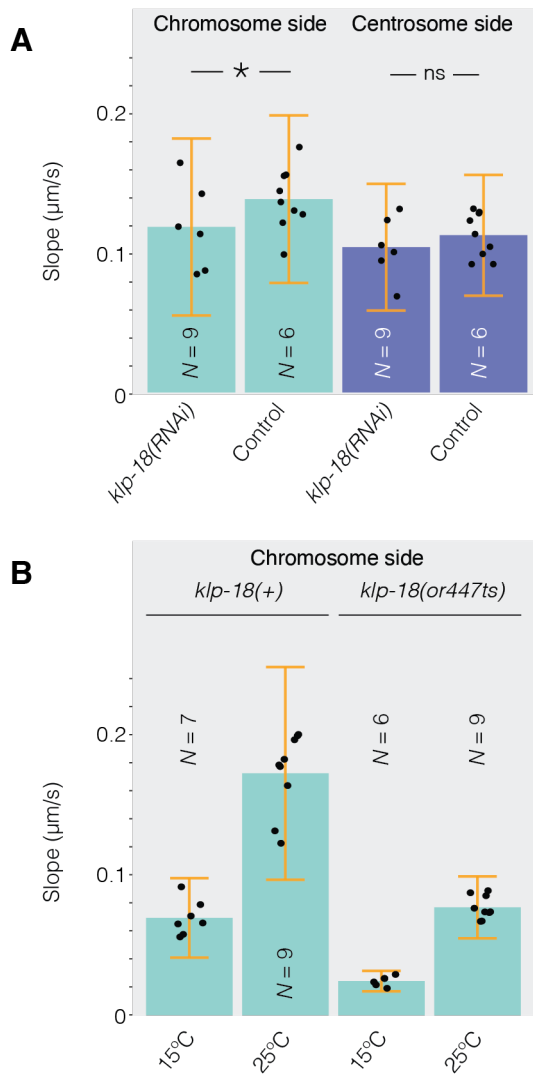

**Figure 4:** (**A**) Front velocities by segmenting the bleached region of $N = 9$ hypomorphic *klp-18(RNAi)* embryos compared to $N = 6$ non-treated embryos. (**B**) Front velocities on the chromosome side by segmenting the bleached region of *klp-18(or447ts)* embryos at ($N = 6$) permissive temperature 15°C and ($N = 9$) restrictive temperature 25°C; corresponding control embryos klp-18(+) at ($N = 7$) permissive temperature 15°C and ($N = 9$) restrictive temperature 25°C. Corresponding velocities on the centrosome side are in Fig S12. Microtubules were labelled using GFP::TBB-2$^{\beta\text{-tubulin}}$. Black dots represent averages of $N$-1 embryos, leaving out, in turn, each embryo (Methods § Statistics on kymograph front slopes). Bars correspond to means; errors are estimated using Jackknife resampling. Light blue bars are values on the chromosome side and dark blue on the centrosome side.

# DISCUSSION

We investigated the dynamics of the microtubules within the metaphase spindle in the nematode one-cell embryo. While we confirmed the lack of global poleward flux by FRAP and photoconversion experiments, we showed that diffusion of tubulin dimers from the cytoplasm could not account for the recovery. Instead, we observed two opposing fronts closing the bleached region. From imaging microtubule plus-ends, we confirmed the fast growth of the spindle microtubules (Fig S13). Modelling such behaviour shows that it causes a recovery of the bleached region uniform along the spindle. In contrast, the front on the chromosome side appears to reflect a distinct mechanism superimposed on the mechanism at work on the centrosome side. We attribute this addition to the kinetochore microtubules sliding along the spindle microtubules, the latter being fixed. Reasons are threefold: (i) While far from the chromosomes, one only sees the spindle microtubules, solely undergoing dynamic instability, and the velocities of both fronts are similar. In contrast, kinetochore microtubules are the majority over the other spindle microtubules only when looking close to kinetochores, leading to differences in the front velocities between both edges. It is plainly consistent with the lack of flux detection in previous works (Labbe et al., 2004; Redemann et al., 2017). (ii) A growth of the microtubules, while attached to the kinetochores, is needed to observe the chromosome side recovery, as found by partially depleting NDC-80, CLS-2$^{CLASP}$ (considering the plateau) or ZYG-9$^{XMAP215}$. It is noteworthy that we also found accelerated flux upon *ndc-80(RNAi)*, *cls-2(RNAi)* and *ska-1(RNAi)*, as expected for poleward flux in other organisms (Matos et al., 2009). (iii) Furthermore, the chromosome-side velocity depends on the status of the attachment of the microtubule at the kinetochore. We probed it by looking over prometaphase and metaphase until early anaphase in unperturbed conditions. We also caused either a premature or a lack of securing of attachment by the SKA complex. In light of the flux measurement close to the chromosomes, our suggested sliding mechanism contrasts with the previous proposal by Redemann and colleagues. Indeed, we found no indication that the kMTs are detaching from the centrosomes and moving towards the kinetochore. However, we cannot exclude that the kinetochore microtubules depolymerise from their minus-ends, consistent with 38% of the ends opened in metaphase (Redemann et al., 2017).

Which mechanism could account for this sliding? A classic mechanism is treadmilling, requiring depolymerisation at the minus end. This mechanism usually involves MCAK as the main microtubule depolymeriser (Arpag et al., 2020; Brust-Mascher et al., 2004). In our case, because the depletion of the homolog of MCAK, KLP-7, does not decrease the recovery velocity, instead tending to increase it, we consider this mechanism unlikely. However, it is not impossible that other microtubule depolymerisers, like KLP-13, katanin, stathmin, and fidgetin, may contribute (Zhang et al., 2007). Notably, the depletion of KLP-13 has a reduced phenotype in the zygote (Kamath et al., 2003; Sonnichsen et al., 2005). Katanin was reported to be inactive during mitosis (Clandinin and Mains, 1993). Stathmin and fidgetin display a weak phenotype upon depletion in the one-cell embryo (Kamath et al., 2003; Lacroix et al., 2014; Lacroix et al., 2016; Sonnichsen et al., 2005). We also considered the transport of the kMTs by cross-linking to sMTs unlikely as these latter are not undergoing poleward flux. We can, therefore, exclude mechanisms involving bridging microtubules (Jagric et al., 2021) or sliding of the overlapping (Miyamoto et al., 2004; Steblyanko et al., 2020; Uteng et al., 2008). We rather propose that the sliding of kMTs over the sMTs is achieved by the kinesin-12 KLP-18. This motor contributes to focusing and organising the meiotic-spindle microtubules (Cavin-Meza et al., 2022; Wolff et al., 2022). Its mammal counterpart, KIF15, was shown to slide parallel microtubules *in vitro* (Drechsler and McAinsh, 2016). Some redundant sliding mechanisms may also exist, such as the transport of the kMT minus ends to the poles by dynein using LIN-5$^{NuMA}$ (Elting et al., 2014). In all cases, the sliding mechanisms must be coupled with the polymerisation of the microtubules at the kinetochore, like in other organisms, to maintain chromosome attachment (Barisic et al., 2021).

We wondered whether restricting the poleward flux to the kMTs is a peculiarity of nematodes. Metaphase is very brief in the nematode embryo, which prevents observing the spindle in a steady state. Furthermore, careful observation of the pole-to-pole distance during the metaphase shows a slow but constant length increase. This situation contrasts with the *Xenopus laevis* spindle prepared from extracts and at steady-state, where a gelification transition was observed (Dalton et al., 2022). A plausible cause of this difference is the strong pulling forces exerted on the spindle poles by the cortical force generators (Grill et al., 2001; Grill et al., 2003; Labbe et al., 2004). Compared to other cells, these forces are very high and participate in spindle elongation. One can speculate that sliding kMTs along the sMTs may mechanically insulate the kinetochores from the poles. In early metaphase, when pulling is weak, the sliding mechanism could generate enough force to ensure a correct function of the spindle assembly checkpoint and subsequent correcting of chromosome mis-attachments. Oppositely, in late metaphase, the strong pulling could be tempered down by letting the kMTs slide along the sMTs, leading to slow spindle elongation. Indeed, while the spindle slowly elongates in the non-treated embryos at that stage, decreasing cortical forces by depleting GPR-1 and GPR-2 reduces this phenotype (Cheeseman et al., 2005; Lewellyn et al., 2010). Overall, such a mechanism may maintain some tension at the kinetochores and be compatible with the SAC and correction mechanism function. Another important difference is the centromere organisation. Nematode chromosomes are holocentric; therefore, the kMTs do not assemble to form a K-fiber. However, a recent electron microscopy study suggests that enough space exists between microtubules in HeLa cells to accommodate molecular motors, making possible a mechanism similar to the one reported here (Kiewisz et al., 2022). Finally, it was recently reported in human cells that the dynamics of microtubules along the spindle axis are not uniform (Conway et al., 2022; Risteski et al., 2022). Furthermore, the minus-ends of the kMTs would also move towards the poles in a poleward flux. Similarly, half of the kMTs in HeLa cells do not reach the corresponding centrosomes (Kiewisz et al., 2022). This observation also suggests that a mechanism similar to the one proposed for the nematode may occur in mammalian cells.

# MATERIAL AND METHODS

## C. *elegans* strains

All *C. elegans* strains were maintained at 20°C and cultured using standard procedures (Brenner, 1974). *C. elegans* worms were grown on NGM plates seeded with OP50 *E. coli* strain. *C. elegans* worm strains and genotypes are listed in Table 1.

| Strain | Genotype | Crossing | Origin and reference |
|---|---|---|---|
| AZ244 | unc-119(ed3) III; ruIs57[pie-1::GFP::tbb-2 +unc-119(+)] V | | CGC (Praitis et al., 2001) |
| EU3068 | ebp-2(or1954[ebp-2::mKate2]) II; ruIs57[pie-1::GFP::tbb-2 +unc-119(+)] V | | CGC (Sugioka et al., 2018) |
| JEP120 | ruIs57[pie-1::GFP::tbb-2 +unc-119(+)] V; gpr-2(ok1179) III. | AZ244 X TH291 | TH291 is a 10x backcross of RB1150 (C. elegans Deletion Mutant Consortium, 2012) |
| JEP92 | unc-119(ed3) ? III; ddIs180[WRM062cF05 spd-2:: 2xTY1 GFP FRT 3xFlag;unc-119(+)] ; ltIs5 [pIC36; pie-1/GFP-TEV-STag::kbp-1; unc-119 (+)] | OD9 x TH231 | OD9: CGC (Cheeseman et al., 2004) |
| | | | TH231: (Decker et al., 2011) |
| JEP97 | unc-119(ed3) ? III; ltIs37 [pAA64; pie-1/mCHERRY::his-58; unc-119 (+)]IV;ddIs44[WRM0614cB02 GLCherry::tbg-1;unc-119(+)] | OD56 x TH169 | OD56: CGC (McNally et al., 2006) |
| | | | TH169: (Woodruff et al., 2015) |

| JEP166 | ddIs44[WRM0614cB02 GLCherry::tbg-1;unc-119(+)]; pJD734_pSW077_Mos1_Pmex-5_mEOS3.2-tbb2 | JEP97 x JDU562 | JDU562: (Dias Maia Henriques, 2018) |
|---|---|---|---|
| JEP172 | ebp-2(gk756) II; ruIs57[pie-1::GFP::tbb-2 +unc-119(+)] V | AZ244 x VC1614 | VC1614: CGC (C. elegans Deletion Mutant Consortium, 2012) |
| SMW36 | klp-18(or447ts), dpy-20 IV; ltSi220[pOD1249/pSW077; Pmex-5::GFP::tbb-2::operon_linker::mCherry::his-11; cb-unc-119(+)] I | | (Wolff et al., 2022) |
| OD868 | ltSi220[pOD1249/pSW077; Pmex-5::GFP-tbb-2-operon-linker-mCherry-his-11; cb-unc119(+)] I. | | (Hollis et al., 2020) |
| JEP176 | ltSi120[[pDC170;Pndc-80:NDC-80 reencoded; cb-unc-119(+)]II; ruIs57[pie-1::GFP::tbb-2 +unc-119(+)] V | AZ244 x OD611 | |
| OD611 | unc-119(ed3) III; ltSi120[[pDC170;Pndc-80:ndc-80 reencoded; cb-unc-119(+)]II #3 | | (Cheerambathur et al., 2017) |
| JEP177 | ltSi122[pDC178;Pndc-80:NDC-80 (8,18,44,51AAAA) reencoded; cb-unc-119(+)]II; ruIs57[pie-1::GFP::tbb-2 +unc-119(+)] V | AZ244 x OD634 | |
| OD634 | unc-119(ed3)III; ltSi122[pDC178;Pndc-80:NDC-80 (8,18,44,51AAAA) reencoded; cb-unc-119(+)]II #1 | | (Cheerambathur et al., 2013) |
| JEP186 | ebp-2(or1954[ebp-2::mKate2]) II unc-119(ed3)III; ltSi462[pNH95; Pmex-5::klp-18::GFP::tbb-2 3'UTR; cb- unc-119(+)]I | EU3068 x OD1495 | EU3068: CGC (Sugioka et al., 2018) OD1495: (Hattersley et al., 2016) |

**Table 1**: Worm strains used in this article and their genotypes.

## RNAi treatment

RNAi was performed by feeding as described previously (Kamath and Ahringer, 2003; Timmons and Fire, 1998). The control embryos for the RNAi experiments were fed with bacteria carrying the empty plasmid L4440. We targeted *ndc-80* by RNAi using the same sequence as in (Cheerambathur et al., 2017) after cloning it into a vector. All other clones came from the Ahringer-Source BioScience library (Kamath et al., 2003). RNAi target, clone ID or sequence used, and feeding conditions are listed in Table 2. We set the RNAi treatment duration so that we did not notice any phenotype suggesting impaired meiosis.

| Target | RNAi clone ID (Ahringer-Source Bioscience) or sequence used | IPTG concentration (mM) | Feeding time (hours) | Incubation temperature (°C) |
|---|---|---|---|---|
| cls-2 | III-4J10 | 4 | 48 | 20 |
| klp-7 | III-5B24 | 4 | 48 | 25 |
| zyg-9 | II-6M11 | 4 | 8 | 25 |
| ndc-80 | Oligo 1: 5' – GGGGACAAGTTTGTACAAAAAAGCAGGCTTGGA TGACAAGTACATTCAGAGATTATACAAATGATC – 3' | 4 | 96 | 20 (Fig 2, S7) |
| | Oligo 2: 5' – GGGGACCACTTTGTACAAGAAAGCTGGGTGGTG GTTCAAGATTCATTTGAATATTAAGTCCACTG – 3' | | | 18 (Fig 3, S9) |
| ska-1 | I-5A22 | 4 | 96 | 18 |
| klp-18 | IV3O14 | 4 | 40 | 18 |
| ani-2 | II-4P19 | 2 | 48 | 25 |
| C27D9.1 | II-4A13 | 4 | 48 | 25 |

**Table 2**: RNAi target, clone ID and treatment conditions used in this article.

## Microscopy: acquisition conditions

### Preparation of C. elegans samples for imaging

*C. elegans* hermaphrodite adults were dissected in M9 medium (50 mM Na2HPO4, 17 mM KH2POone, 86 mM NaCl, 1 mM MgSO4) (Brenner, 1974). The released embryos were deposited on an agarose pad (2% w/v agarose, 0.6% w/v NaCl, 4% w/v sucrose) between slide and coverslip or between two 24x60 cm coverslips. To confirm the absence of phototoxicity and photodamage, we checked for normal rates of subsequent divisions (Riddle, 1997; Tinevez et al., 2012).

### Imaging condition for photobleaching and photoconversion experiments

Embryo imaging to perform FRAP was achieved on a scanning confocal microscope with deconvolution (LSM 880 airyscan, Zeiss) using an apochromatic 63x/NA1.4 oil objective. Acquisition parameters were controlled by Zen black software. We imaged at 18°C, except otherwise stated, using the CherryTemp temperature control system in all cases (CherryBiotech, Rennes, France). With all strains expressing GFP::TBB-2$^{\beta\text{-tubulin}}$, we performed photobleaching using an argon laser at wavelength 488 nm and 70 µW power for 75 iterations (photobleaching time ~ 1.9 s). We then imaged using the same laser but at a 2.30 µW power. Five images were taken before photobleaching to get the basal fluorescence of the embryo. The laser power was measured at the objective output just before each microscopy session. We set to bleach a 2.6 x 19.5 µm area within the mitotic spindles (the embryo was oriented horizontally) (Fig1A). When accommodating two bleached regions per half-spindle (Fig 2AB), their width read 1.3 µm. Images were acquired at 12.5 Hz.

In the case of photoconversion experiments, we used the strain JEP166 expressing mCherry::TBG-1$^{\gamma\text{-tubulin}}$ and mEOS3.2::TBB-2$^{\beta\text{-tubulin}}$. The mEOS3.2::TBB-2$^{\beta\text{-tubulin}}$ labelling was obtained from the JDU562 strain, a kind gift from Julien Dumont's team. The mCherry::TBG-1$^{\gamma\text{-tubulin}}$ channel enables the images' registration on the centrosomes' position (see § Image processing). Five images were taken before photoconversion to have the basal fluorescence of the embryo. Photoconversion was performed using a 405 nm UV laser at 10 µW for 35 iterations (photoconversion time ~ 0.850 s). We photoconverted a 13 x 1.3 µm area within the mitotic spindle after vertically orienting the anteroposterior axis of the embryo. Images were acquired with a 561 nm laser at 9 µW power and 3 frames per second.

To study the localisation of KLP-18, we imaged *C. elegans* one-cell embryos at the spindle plane, viewing from the nuclear envelope breakdown (NEBD) until late anaphase. We used a Leica DMi8 spinning disk microscope with Adaptive Focus Control (AFC) and an HCX Plan Apo 100x/1.4 NA oil objective. Illumination was performed using a laser with an emission wavelength of 488 nm, and we used a GFP/FITC 4 nm bandpass excitation filter and a Quad Dichroic emission filter. Images were acquired using an ultra-sensitive Roper Evolve EMCCD camera that was controlled by the acquisition device by Inscoper, SAS. During the experiments, the embryos were kept at about 20°C.

In all experiments, the anaphase onset was taken as a time reference. Images were stored using OMERO software (Li et al., 2016).

### Image Processing: Automatic recognition of the bleached region edges and measuring the slopes of the fluorescence-recovery fronts

#### Assembling a homogenous set of embryos

Observing the kymograph (Fig 1B, S3) suggested that the recovery proceeded by two fronts moving inwards. We set out to measure the velocity of these fronts. We measured the laser power at each session to safeguard against intensity variation between sessions. It also enabled us to select a set of embryos bleached in similar conditions. Furthermore, the spindle lies 10-15 µm deep in the sample, further introducing variability. Overall, we considered only the experiments where the FRAP efficacy reached at least 30%, assessed by comparing the fluorescence intensities before and after the FRAP event. We checked that we performed bleaching during metaphase, i.e. between -120 and -30 s before the anaphase onset; it ensured a full fluorescence recovery prior to anaphase onset. For the *ska-1(RNAi)* experiments, this range was reduced to [-60 -30] s to correspond to the time period when SKA-1 is recruited in non-treated embryos (Cheerambathur et al., 2017). Finally, the distance between the metaphasic plate and the bleached region was measured to guarantee that a bright corridor separated the bleached region from the hollow due to the chromosomes, ensuring a robust measure of the chromosomal front speed (Fig 1B, S3).

#### Obtaining a kymograph for each embryo

We performed a rigid registration of the raw images using the slightly modified version of the so-called plugin (publicly available v. 1.4) under Icy (de Chaumont et al., 2012). This registration focused on the centrosome position on the side of the bleached half-spindle, preventing spindle displacement or rotation from contributing to the measured front motion (Fig 1, S1, S2A). We then applied a median filter under Fiji to limit the noise (Schindelin et al., 2012). We computed the kymograph by considering a region of interest of a length of 23.4 µm and a width of 3.12 µm (Fig S2B). To do so, we used Fiji and performed a median projection along a direction transverse to the spindle axis (Fig S2C).

#### Averaging kymographs over several embryos

To ensure that each embryo contributed equally, we performed a histogram equalisation among the individual embryos' kymographs using the histogramMatcher class in sciJava (interfaced using Beanshell) (Rueden et al., 2021). Finally, we registered the kymographs aligning on the centrosome-side edge of the bleached region at the first time after bleach, using a supervised home-designed Matlab plugin (Fig S2C, arrowhead). We then averaged the kymographs pixel-wise (Fig S2D). We then cropped the embryo-averaged kymograph to consider only the 40 first seconds and the bleached half-spindle.

#### Kymograph segmentation

Next, we segmented the bleached area in the kymograph. As a preprocessing, we convolved the kymograph with a 3x3 kernel reading 5 in the central element and 1 elsewhere (Gaussian blur). We then autonomously detected the non-bleached area being conservative and considering only the pixels evidently non-bleached. To do so, we set the background class as the 60% brightest pixels. We then performed grey morphology (to remove micro/dot regions due to noise). Practically, we used two iterations of opening with a 1-pixel (130 nm) circular structural element and two closing iterations with the same element (Fig S2E, black line). Similarly, we set the bleached area as the 25% darkest pixels and performed the grey morphology (Fig S2E, orange line). We used neighbourhood information to decide about the pixels with intermediate grey levels. We trained a machine learning algorithm, namely random forests featuring 100 trees (Breiman, 2001), using the above classes for training and all kymographs in a set acquired under the same condition. We then applied the algorithm to classify the pixels within the kymograph (Fig S2E, red line) using Weka under Fiji (Beanshell script) (Arganda-Carreras et al., 2017). Finally, we performed grey morphology closing (2 pixels circle), watershed and the same closing again. It enabled us to remove artefactual dotty regions while preserving the largest one corresponding to the bleached area. We then selected the largest segmented object. Indeed, because the recovery was never complete (Fig 1B) and the contrast was low, a small artefactual region may appear in the tail of the main one. Finally, we dilated the shape using a 1-pixel-radius circle twice, making the resulting shape convex.

*Kymograph boundary fitting*

Having segmented the bleached area, we fitted a line along the boundaries (between 2 and 30 s) to get the mean front velocities. We also computed the mid-curve (average between the positions of the two edges at each time point) and fitted it with a line.

*Statistics on kymograph front slopes*

We used a resampling method to get the error bars of the front velocities since we could not perform the computation on individual embryos because the images were too noisy. We used the Jackknife method, which requires computing the slope over a set containing all embryos but one (Efron and Tibshirani, 1993; Quenouille, 1956). We thus obtained a standard error on the averaged value estimated considering all embryos and could perform *t*-tests against the null hypothesis that the slope is 0.

## Statistics representation

To compare conditions, we used the Jackknife again (Arvesen, 1969; Schechtman and Wang, 2004) or Pearson correlation when relevant. For the sake of simplicity, we depicted confidence levels using diamonds or stars (***, $P \leq 0.0005$; **, $P \leq 0.005$; *, $P \leq 0.05$; ◇ $P \leq 0.1$; n.s., $P > 0.1$). This latter means non-significant.

## Acknowledgements

The strains TH291, TH231, and TH169 are kind gifts from Prof A. A. Hyman. Dr J. Dumont kindly offered JDU562. We thank Dr Gregoire Michaux for the feeding clone library and technical support. We are also thankful to Dr Arshad Desai for the gift of strains OD868, OD611 and OD634. The strain SMW36 was a kind gift of Prof Sarah M. Wignall. We also thank Drs. Gilliane Maton, Fabrice Mahé, Anne Corlu, Christophe Heligon, Rebecca Smith, Sebastien Huet, Flora Demouchy, Gregoire Michaux, Anne Pacquelet, Stéphanie Dutertre, Xavier Pinson, Marc Tramier, Ostiane d'Augustin for discussions about the project. Some strains were provided by the Caenorhabditis Genetics Center (CGC), funded by the National Institutes of Health Office of Research Infrastructure Programs (P40 OD010440; University of Minnesota). We also acknowledge La Ligue contre le cancer (comites d'Ille-et-Vilaine et du Maine-et-Loire). Microscopy imaging was performed at the Microscopy Rennes Imaging Center (MRIC), UMS 3480 CNRS/US 18 INSERM/University of Rennes 1. MRIC is a member of the national infrastructure France-BioImaging, supported by the French National Research Agency (ANR-10-INBS-04). In particular, we acknowledge the funding of the Zeiss Airyscan confocal microscope by EU funding FEDER under reference CARE Phase 2.

## Author contribution

Conceptualisation: JP, LLM; Data curation: NS, MDS, CT, LC, JP; Formal analysis: NS, MDS, CT, LC, JP; Funding acquisition: JP, HB, LLM; Investigation: NS, MDS, CT, PF, LC, SP, JP; Methodology: NS, MDS, HB, LC, JP; Project administration: JP; Software: NS, JP, HB; Supervision: LLM, JP, LC; Validation: HB, LC, JP; Visualisation: NS, JP; Writing NS, JP; Writing – review and editing: LC, NS, MDS, JP, LLM, HB.

## Conflict of interest

The authors declare that they have no conflict of interest.

# Supplemental information to the article "Kinesin-12 KLP-18 contributes to the kinetochore-microtubule poleward flux during the metaphase of *C. elegans* one-cell embryo." By Soler et al.

## Supplemental Text

We have detailed here complementary investigations and the approaches developed specifically for this work while the details for applying existing methods are reported in the Supplementary Methods section below.

1. ### Cytoplasmic tubulin-dimer diffusion does not account for recovery after photobleaching.

During metaphase, we bleached the fluorescently labelled tubulin in a band-shaped region, 2.6 μm-wide, perpendicular to the spindle axis and covering the spindle but extending up to the cytoplasm (Fig S1A). The proximal boundary of the bleached area was between 0.5 and 2 μm away from the chromosomes (defined as the area devoid of microtubules). Since the spindle is moving and slowly elongating during metaphase, we performed an intensity-based registration of the images over time using the icy software (de Chaumont et al., 2012). We used the centrosome on the bleaching side as a spatial reference (see Methods § Image processing, Fig S1A). We monitored the fluorescence recovery over time and, assuming a single microtubule population, we fitted the time-recovery curve with a single exponential over a 60-second range during metaphase (Giakoumakis et al., 2017; Girao and Maiato, 2020) (Suppl Methods § Imaging conditions). We used a global fitting approach to safeguard against variability between embryos and obtained the confidence interval by thresholding the empirical likelihood (Bouvrais et al., 2021). In doing so, we imposed the same model parameters on each embryo of a given data set and maximised the product of the embryo-wise likelihoods (Beechem, 1992). We found a recovery half-time inside the spindle about three times longer than in the cytoplasm, although the recovery was never complete (Fig S1B). The spindle half-life was consistent with previous measurements (Redemann et al., 2017) and suggested that free fluorescent tubulin was available within seconds within the spindle.

2. ### Bleached-area edge closure depends on microtubule dynamics.

To challenge the link between the observed fronts (Fig 1B) and the microtubule dynamics, we set to decrease the growth rate of microtubules globally. ZYG-9$^{XMAP215}$ was a proper candidate as it affects all microtubules, astral, spindle and the ones attached at the kinetochores (Encalada et al., 2005; Fernandez et al., 2009; Lacroix et al., 2018; Srayko et al., 2005). However, we restricted ourselves to a partial depletion to preserve a functional spindle assembly (Matthews et al., 1998). We observed a delayed rotation of the spindle, confirming the penetrance of the treatment (Bellanger et al., 2007). We then performed the FRAP experiments and measured the absolute value of the slopes of the fronts (Fig S1D, E, replica in S1F). We observed a significant decrease in the slopes on both the chromosome and the centrosome sides. It suggested that microtubule dynamics contributed to the mechanisms accounting for these fronts.

3. ### Modelling recovery on the centrosome side

We reckoned that the front displacement on the centrosome side might be accounted for by the dynamics of the sMTs (growth and shrinkage) combined with diffraction due to imaging. Indeed, the fast recovery after photobleaching within the cytoplasm suggested that non-bleached tubulin diffuses fast and supports the growth of microtubules with bright dimers in the bleached area. We set out to model the kymograph by combining known microtubule dynamics and simulating the diffraction due to the microscope.

**Spindle- and kinetochore-microtubule dynamics**

We considered two microtubule populations within the spindle: one emanating from the centrosome and one attached to the kinetochore. In doing so, our analysis ignored the long-lived microtubules. These microtubules could be the few kMTs directly connecting the centrosome to the kinetochores, but their proportion is low (Redemann et al., 2017). Long-lived microtubules could also correspond to overlapping sMT. However, these latter were mostly expected around the metaphasic plate as it was suggested that the microtubules could hardly perforate the chromosome domain (Redemann et al., 2017; Schneider et al., 2022). The region of interest used in extracting the experimental kymograph did not include many of these, particularly when close to the kinetochores. We thus ignored long-lived microtubules in modelling and focused on the first 30 seconds after photobleaching.

We modelled the spindle microtubules (sMTs) as free-growing microtubules until reaching the chromosomes (Zelinski et al., 2012). Microtubule dynamics parameters were obtained from the literature or our measurements (Suppl Table S1). It resulted in an exponential-like decay of the microtubule density from the centrosome to the kinetochores, consistent with electronic microscopy imaging (Fig 5A, red line) (Redemann et al., 2017). In particular, since the growing rate is critical to our model, we measured it to $0.98 \pm 0.02$ µm/s (average and standard deviation of the Gaussian) in our conditions (Suppl Text §4, Fig S13). At the kinetochores or chromosomes, modelled as impassable obstacles (Schneider et al., 2022), we set the catastrophe rate to 1 per second. Indeed, varying by one order of magnitude around this value suggested that the density profile of the sMTs was poorly dependent on this parameter. Furthermore, such a rate corresponded to a 1 s residency time before detaching from the kinetochore and was similar to observation on astral microtubules at the cell cortex (Bouvrais et al., 2021). This catastrophe rate was about 5 times higher than at free microtubule ends. We adjusted the spindle length as the distance between the two fluorescent centrosomes on the images. We computed it using the intensity profile of non-treated embryos, averaged during 0.5 second, just after photobleaching. Technically, we smoothened the intensity profile using a cubic spline and set the position of the peaks (assumed to correspond to the centrosomes) as the zero-crossings of the derivative.

We next modelled the kinetochore microtubules. We assumed that the microtubules were permanently growing at the kinetochores due to the tension, as reported in other organisms (Cheeseman et al., 2004; Suzuki et al., 2015). The corresponding speed was set close to our observation, 0.1 µm/s. Doing so, we obtained kMT minus ends uniformly distributed along the spindle, as expected (Fig S5E, black line) (Redemann et al., 2017). We set the ratio of kMT to sMT to 1.5 at 1 µm from the kinetochore after the electron micrograph (Fig S5E).

**Recovery after Photobleaching**

We then modelled the fluorescence recovery after photobleaching. We assumed no exchange of tubulin along the lattice, i.e. no repair of microtubule defects (Schaedel et al., 2019). Recovery for the sMTs was accounted for by the dynamic instability of the microtubules. Technically, we modelled the bleaching as an entire disappearance of fluorescence in the corresponding region. About recovery, based on the fast diffusion of tubulin dimers in the cytoplasm, we assumed that fluorescent tubulin dimers are readily available and modelled the growth of bleached microtubules with fluorescent subunits. It was equivalent to scaling the fluorescence curve in the bleached region with a coefficient (Fig S5A). This coefficient increased from 0 at bleaching time to 1 when reaching the total recovery. We modelled the evolution of this coefficient using a single exponential function with a characteristic time of 30.4 s, as measured in this study. In doing so, we assumed that only free diffusion was involved. In the region considered experimentally, the sMTs were in the majority in most places. Recovery on the kinetochore side was again attributed to the dynamic instability mechanism but with kMTs poleward flux superimposed (Fig S5A). The flux rate was adjusted so that the edge of the simulated bleached region velocity mimicked the experimental value in non-treated embryos. Finally, we formed a pseudo kymograph (Fig S5D).

**Diffraction due to microscope imaging**

We modelled the diffraction by a Gaussian PSF, whose standard deviation was obtained by fitting the experimentally measured PSF and read 149 nm. To do so, we imaged fluorescent beads of diameters 175 nm (PS-Speck Microscope Point Source) on the same microscope. We fitted the experimental image with a Gaussian (Kirshner et al., 2013) to obtain the PSF.

We then convolved such a Gaussian, in 1D, with the simulated density of the microtubules at each time (Fig S5BC) and obtained a convolved kymograph mimicking the experimental kymograph (Fig S4A compared to Fig 1B and S5D).

We then convolved such a Gaussian, in 1D, with the simulated density of the microtubules at each time to get fluorescence profiles that accounted for the effect of microscopy imaging (Fig S5BC). Last, we obtained a convolved kymograph mimicking the experimental kymograph (Fig S5A compared to Fig 1B). Notably, it differed from the one in Fig S5D, not accounting for microscopy diffraction.

**Segmentation of the pseudokymograph and velocity extracting**

We analysed this simulated kymograph (pseudokymograph) similarly to the experimental one to extract the slopes of the fronts (Methods § Kymograph boundaries fitting). One parameter was changed: we used a threshold of 13% on the intensity histogram to segment the bleached area. Indeed, the absolute intensity in the simulation differed from the experimental one. This first segmented region, the core of the bleached region, was used to train machine learning. We then performed the further steps of the segmentation and kymograph edge fitting as for experimental kymographs. We obtained the front velocity of 0.016 μm/s on the centrosome side, in the same order of magnitude as the experiments. The bare growth and shrinking of the microtubules emanating from the centrosome (Fig S5A) combined with the diffraction due to microscope imaging was enough to lead to an apparent front moving. It could account for the recovery seen on the centrosome side despite the sMTs not undergoing flux (Fig S5B). When no kMT is added, the front motions, without flux, were predicted to be almost identical on both sides. Adding fluxing kinetochore microtubules broke this symmetry in agreement with experiments (Fig S5C).

| Quantity | Value | Reference |
| --- | --- | --- |
| MT growth rate | 0.65 μm/s | (Srayko et al., 2005) |
| MT shrinking rate | 0.84 μm/s | (Kozlowski et al., 2007) |
| Catastrophe rate | 0.28 /s | (Lacroix et al., 2018) |
| Rescue rate | 0.44 /s | (Lacroix et al., 2018) |
| Catastrophe against Chromosome | 1 /s | (Bouvrais et al., 2021) |
| Spindle length | 14.5 μm | This study |
| kMT poleward displacement | 0.1 μm/s | This study |

| sMT recovery characteristic time | 30.4 s | This study |
|---|---|---|
| Bleached region boundary from the centrosome | 3.18 μm | This study |
| Bleached region width | 2.6 μm | This study |
| PSF standard deviation | 149 nm | This study |

**Supplemental Table S1**: Values used in the simulation and associated references.

**Validating the model**

We set out to challenge the proposed model experimentally. We predicted that the slope on the centrosome side scaled inversely with the length of the spindle (Fig S4C). We tested the correlation between the spindle length and the front velocity on the centrosome side in most of the experiments presented in this article (Fig S4B). We observed a clear anti-correlation (Pearson $r = -0.66$, $p = 0.001$) as predicted by the model (Fig S4C). It was indicative of a global regulation since such an anti-correlation was seen across variegated depletions or mutation, including the microtubule dynamics either overall (ZYG-9, e.g.), either at the centrosome (KLP-7) or kinetochore (NDC-80, CLS-2). Since microtubule dynamics caused the front on the centrosome side, finding it dependent on spindle length was consistent (Fu et al., 2015; Maffini et al., 2009; Nehlig et al., 2021). Interestingly, such an anti-correlation was not seen experimentally when looking at the front velocity on the chromosome side (Fig S6). It is consistent with the model prediction (Fig S4C). We concluded that the front motion closing the bleached area on the centrosome side was likely not caused by any microtubule flux. Instead, it was due to the combination of spindle microtubule dynamics and microscopy imaging.

### 4. Measuring the growing dynamics of spindle microtubules

We measured the displacement of the microtubule plus-ends within the spindle using EBP-2$^{EB1}$ labelling (Srayko et al., 2005). We imaged the doubly labelled strain EBP-2$^{EB1}$::mKate2; GFP::TBB-2$^{\beta\text{-tubulin}}$ at two frames per second to keep the SNR high. We registered these images over time to suppress the effect of spindle displacement as described above, then denoised the images with the Kalman filter (Kalman, 1960) (Fig S13A) and formed the kymograph as previously done (Fig S13B). We then measured the microtubule plus-ends velocity using the directionality ImageJ plugin within Fiji (Suppl Methods § Measuring the growth rate of microtubule plus ends), disregarding whether this velocity was due to microtubule growth or displacement (Liu, 1991; Schindelin et al., 2012; Schneider et al., 2012). We noticed that the distribution of comet orientation angles for a single embryo corresponded to a single Gaussian (S13D). We thus performed a global fit of individual embryos transforming angles into velocities (equation in Methods § Measuring the growth rate of microtubule plus ends) and found a growth rate of $0.98 \pm 0.02$ μm/s (average and standard deviation of the Gaussian), consistent with previous measurements (Lacroix et al., 2018; Srayko et al., 2005). The histogram of EBP-2$^{EB1}$ comet speed is reproduced in Fig S13C. It suggested that a single dynamical behaviour was present far from the chromosomes (where we measured), likely corresponding to the growing sMTs.

### 5. Front velocities did not depend on EBP-2

EBP-2$^{EB1}$ is a member of a minimal treadmilling system (Arpag et al., 2020). We thus performed a similar FRAP experiment and analysis under null mutant condition *ebp-2(gk756)* (Fig S11). We found a mild reduction of the recovery-front rate on the chromosome side, consistent with the lack of strong phenotype previously reported in the embryo for EBP-2 depletion (Kamath et al., 2003; Sonnichsen et al., 2005). However, several proteins involved in the kinetochore may be impacted by EBP-2 depletion, like SKA-1 (Lange et al., 2019) or dynactin subunit DNC-2$^{p-50}$

(Barbosa et al., 2017). Overall, we concluded that the recovery front on the chromosome side was due to a mechanism likely independent of EBP-2$^{EB1}$.

# Supplemental Methods

## *Imaging condition for measuring the velocity of microtubule plus-ends*

Embryo imaging was performed on a scanning confocal microscope with deconvolution (LSM 880 airyscan, Zeiss) using an apochromatic 63x/NA1.4 oil objective. Acquisition parameters were controlled by Zen black software. Imaging temperature was controlled using the CherryTemp temperature control system (CherryBiotech, Rennes, France). We also used focus maintaining for this particular experiment to compensate for drift. We measured the comets' velocity of microtubule plus-end to assess the microtubule growth rate. We used the strain EU3068 expressing EBP-2$^{EB1}$::mKate2 GFP::TBB-2$^{β-tubulin}$ (Sugioka et al., 2018). Image acquisition was performed with a HeNe laser at wavelength 594 nm and 3.35 µW power. The laser power was measured at the objective output at the beginning of each microscopy session. Images were acquired at 2 Hz on a single plane.

## *Measuring the growth rate of microtubule plus ends*

We measured the microtubule's growth rate within the spindle by looking at their plus-end displacement. First, we registered the images using the GFP::TBB-2$^{β-tubulin}$ channel to keep the posterior centrosome immobile in the image stack as described above (Methods § Averaging kymograph over several embryos) (Fig S13A). Indeed, our study of the flux focused on the spindle half. Using the Kalman filter, we denoised the EBP-2$^{EB1}$::mKate2 channel (Kalman, 1960). The EBP-2$^{EB1}$ labelled plus-end displacement appeared as an oblique line on the kymograph (Fig S13B). This latter was obtained by considering a region of interest of a length of 4.68 µm and a width of 3.12 µm centred on the posterior half-spindle. We formed the kymograph as previously described. We measured the direction and slope of these lines using the slightly modified version of the directionality plugin on ImageJ ((Tinevez et al., 2017), https://imagej.net/plugins/directionality). It produced a histogram counting the lines per direction (Fig S13D). We modelled the orientation distribution with a Gaussian. $a_d . e^{-(x-\mu_d)^2 / 2\sigma_d^2}$ where normalisationentation, $\mu_d$ the average orientation, $\sigma_d$ the standard deviation and $a_d$ the normalisation factor. We computed the velocity corresponding to the direction distribution for each embryo. However, the focus-maintaining system caused variable frame rates. We thus performed a global fit over all embryos, sharing the parameters to safeguard embryo-to-embryo variability again (Beechem, 1992). Practically, we used the model $\hat{a}.\exp\left(-(\text{atan}(v.\frac{\tau}{\rho}) - \text{atan}(\hat{\mu}.\frac{\tau}{\rho}))^2 / 2\text{atan}(\hat{\sigma}.\frac{\tau}{\rho})^2\right)$ with $\hat{a}$ the fit-estimated normalisation factor, $v$ the velocity, $\tau$ the cycle time at imaging (different for each embryo), $\rho$ the pixel size (resolution), $\hat{\mu}$ the estimated velocity and $\hat{\sigma}$ the standard deviation (Fig S13C).

# Supplemental Figures

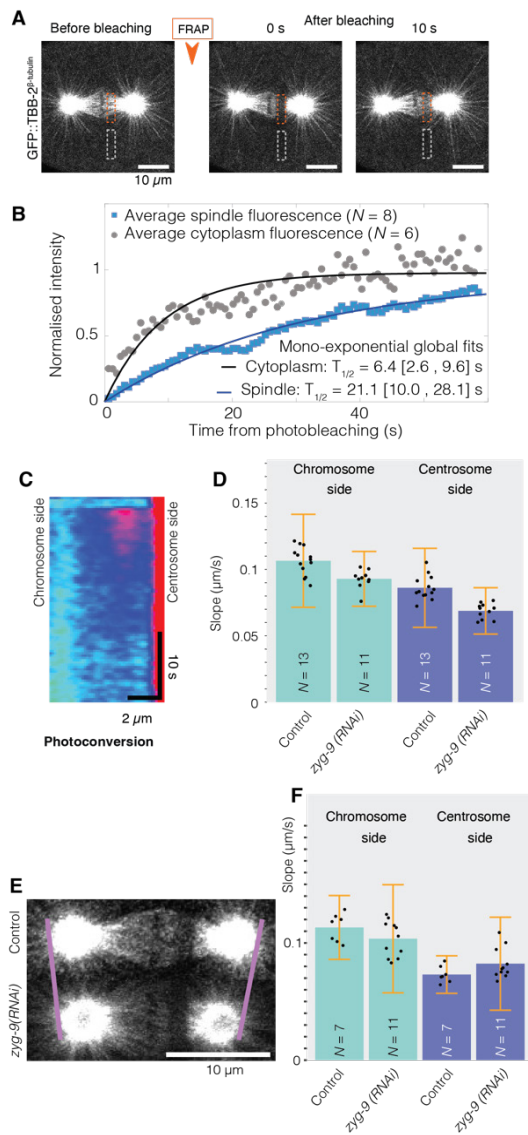

**Figure S1: Tubulin fluorescence recovery after photobleaching and photoconversion in the metaphasic spindle of *C. elegans*.** (**A**) Representative confocal live images before and after photobleaching (time above the stills). The posterior centrosome is on the right-hand side. Microtubules are labelled with GFP::TBB-2$^{\beta\text{-tubulin}}$. The scale bar corresponds to 10 μm. Orange and grey dashed boxes depict the regions used to assess fluorescence recovery in the spindle and cytoplasm, respectively. (**B**) Fluorescence recovery averaged over GFP::TBB-2$^{\beta\text{-tubulin}}$ non-treated embryos bleached: (blue) $N = 8$ in the spindle and (grey) $N = 6$ in the cytoplasm. Lines correspond to the global fit by a single exponential model (Giakoumakis et al., 2017) (Suppl Text §1). The corresponding half-lives are reported with the 95% confidence interval within brackets. (**C**) kymographs after photoconversion within the mitotic spindle averaged over $N = 7$ mEOS3.2::TBB-2$^{\beta\text{-tubulin}}$ with mCherry::TBG-1$^{\gamma\text{-tubulin}}$ embryos. The colour scale ranges from blue for dark pixels to red for bright areas. The centrosome was located on the right-hand side. (**D**) Front velocities by segmenting the bleached region of $N = 11$ *zyg-9(RNAi)* treated embryos and $N = 13$ control ones. Black dots represent the averages of $N$-1 embryos, leaving out, in turn, each embryo (Methods § Statistics on kymograph front slopes). Bars correspond to means; error bars are estimated standard errors using Jackknife resampling. Light blue bars are values on the chromosome side and dark blue on the centrosome side. (**E**) Exemplar micrographs of single GFP::TBB-2$^{\beta\text{-tubulin}}$ embryo used for FRAP experiment and submitted to (bottom) *zyg-9(RNAi)* or (top) corresponding control treatments. (**F**) Replica of the experiment (D) using $N = 11$ *zyg-9(RNAi)* treated embryos and $N = 7$ control ones. Light blue bars are velocity values measured on the chromosome side and dark blue on the centrosome side. In D-F, we used strains with labelled microtubules GFP::TBB-2$^{\beta\text{-tubulin}}$.

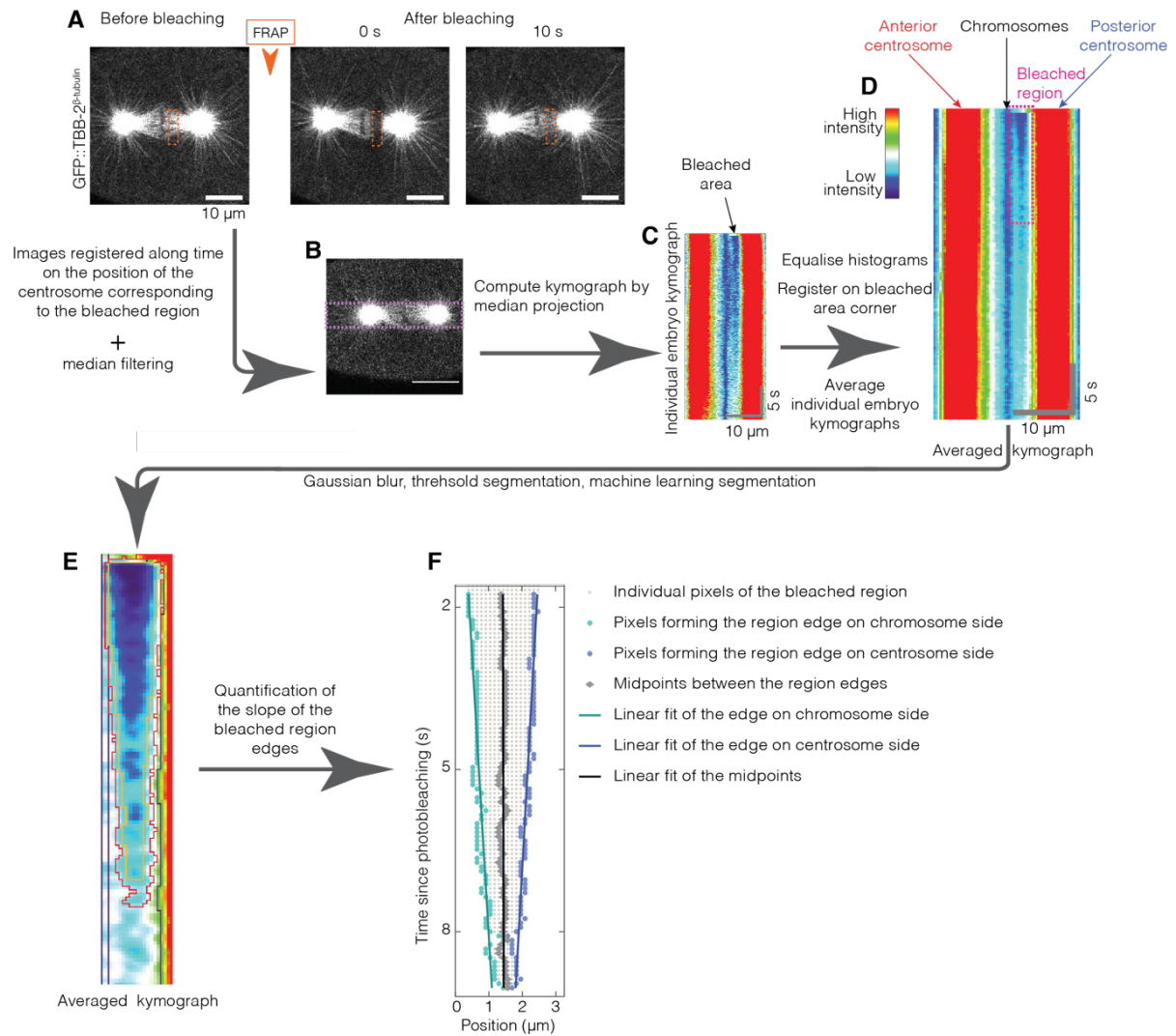

**Figure S2: Flow diagram of the image processing measuring front velocities** (**A**) Representative confocal live images before and after photobleaching. GFP::TBB-2$^{\beta\text{-tubulin}}$ labels microtubules. The scale bar corresponds to 10 μm. The orange dashed box depicts the photobleached region. (**B**) Images were registered in time on the centrosome on the same side as the bleached region, and (**C**) a kymograph was formed by median projection along the vertical dimension of the mauve dashed box superimposed to the micrograph (B). (**D**) Kymographs of the individual embryos were histogram-equalised and averaged. (**E**) Typical averaged kymograph after Gaussian blur used for segmentation. We first used a thresholding method to determine (black contour, outside) the background pixels and (orange contour, inside) the bleached-region pixels. (red contour) We then propagated this segmentation with a random-forests machine-learning classification of the pixels (*weka*) and obtained the segmented region, post-processed using grey morphology. (**F**) This region is contoured by light blue points on the chromosome side and dark blue ones on the centrosome side, while the computed middle at each time is depicted with grey dots. We finally linearly fitted the boundary of this region and the mid-line.

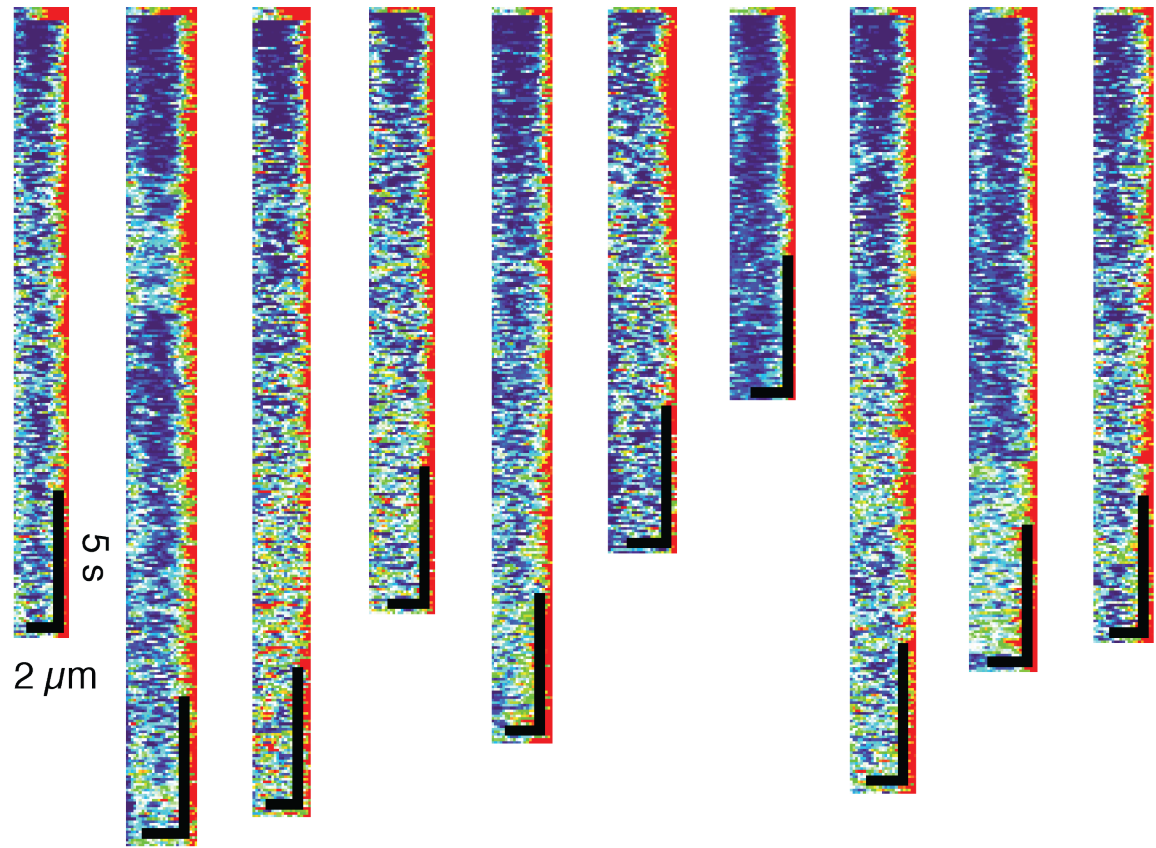

**Figure S3: Individual embryo kymographs** of the posterior half of the spindle reported for $N = 10$ GFP::TBB-2$^{\beta\text{-tubulin}}$ non-treated embryos, with centrosome on the right-hand side. They were obtained from time-registered images (Methods § Image processing). Scale bars correspond to 2 μm horizontally and 5 s vertically. These embryos are the same as in Fig 1. The colour scale ranges from blue for dark pixels to red for bright areas.

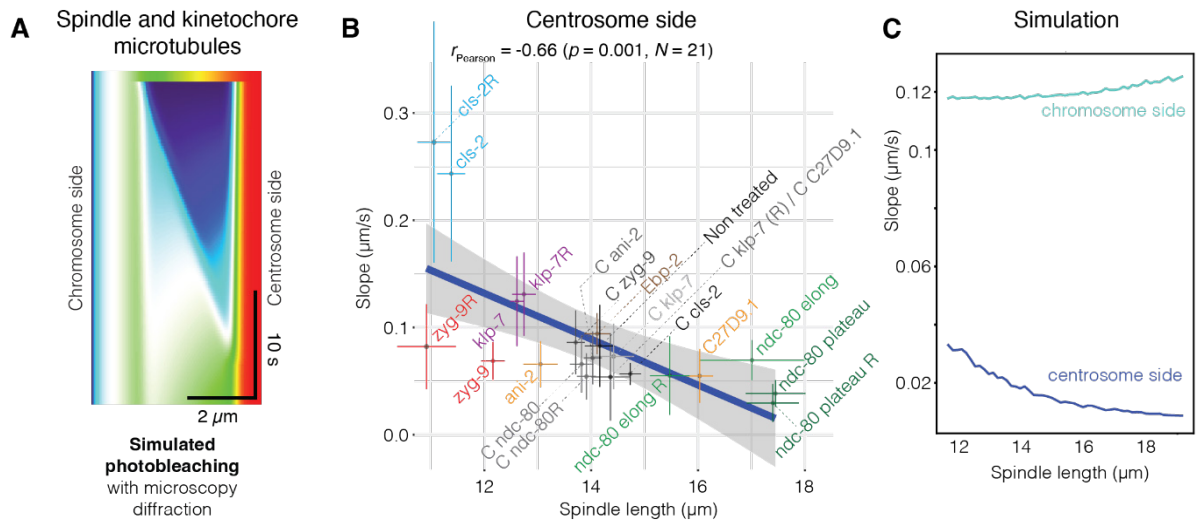

**Figure S4: Front velocity on the centrosome side did not require flux but depended on spindle length.** (**A**) Simulated kymograph including not-fluxing sMTs and kMTs submitted to poleward flux (Suppl Text §3). It accounts for diffraction by microscopy imaging. Corresponding simulated slopes read 0.12 µm/s on the chromosome side and 0.016 µm/s on the centrosome side. The colour scale ranges from blue for dark pixels to red for bright areas. Simulation parameters are reported in Suppl Table S1. (**B**) The experimental front velocities on the centrosome side were anti-correlated with the spindle lengths. The formers were measured in various conditions (indicated). We used strains with labelled microtubules GFP::TBB-2$^{\beta\text{-tubulin}}$. Spindle length was measured at the time of bleaching. The conditions are detailed in other figures and the Methods. Suffix R stands for the replica experiment, and C for the control experiment attached to the mentioned treatment. Pearson correlation coefficient and the corresponding test are indicated above the plot. Bars correspond to means; error bars are standard errors. The grey-shaded region corresponds to the confidence interval at 0.95 on the line coefficients. (**C**) The model predicted an anti-correlation between (blue) the centrosome-side front slope and the spindle length. (turquoise) In contrast, no correlation was predicted on the chromosome side, matching the experimental observation (Fig S6).

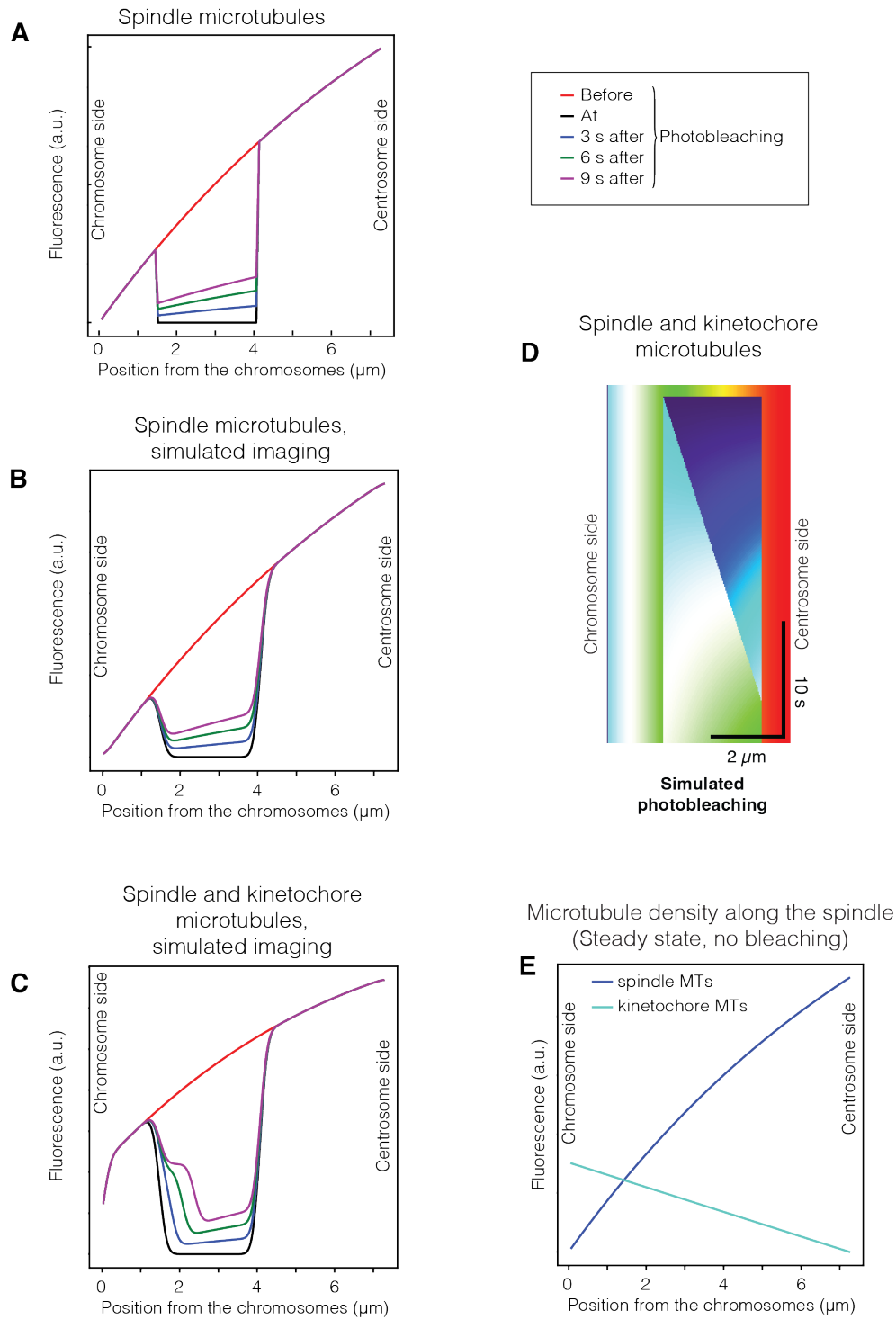

**Figure S5: Modelling the fluorescence recovery and the front displacement. (ABC)** Fluorescence profiles along the spindle (coloured lines) were modelled at various times after simulated bleaching, considering (AB) only the microtubules emanating from the centrosomes (sMTs) or (C) also the kinetochore microtubules (kMTs). The effect of microscopy imaging was (A) ignored or (BC) accounted for by convolution with the point spread function (Suppl Text §3). The portion of the curve within the bleached region was scaled by a coefficient so that the bleached fraction of tubulin underwent an exponential decay with time. **(D)** Simulated kymograph including both not-fluxing sMTs and kMTs submitted to poleward flux. We have not accounted for microscopy-imaging diffraction compared to Fig S4A done with the same simulation parameters. The colour scale ranges from blue for dark pixels to red for bright areas. **(E)** Density of sMTs and kMTs along the spindle axis, equivalent to their fluorescence in our model. Simulation parameters are reported in Suppl Table S1.

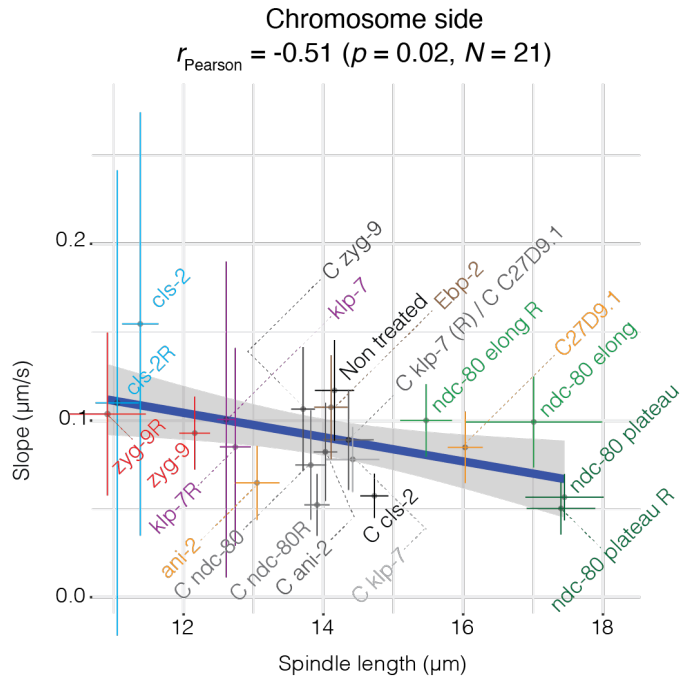

**Figure S6:** The measured front velocities on the chromosome side did not significantly correlate with the spindle length. The formers were measured in various conditions (indicated) by segmenting the bleached region. We used strains with labelled microtubules GFP::TBB-2$^{\beta\text{-tubulin}}$. Spindle length was measured at the time of bleaching. The conditions are detailed in other figures and Methods. Suffix R stands for the replica experiment, and C for the control experiment attached to the mentioned treatment. Pearson correlation coefficient and the corresponding test are indicated above the plot. Bars correspond to means; error bars are standard errors. The grey-shaded region corresponds to the confidence interval at 0.95 on the correlation-line coefficients.

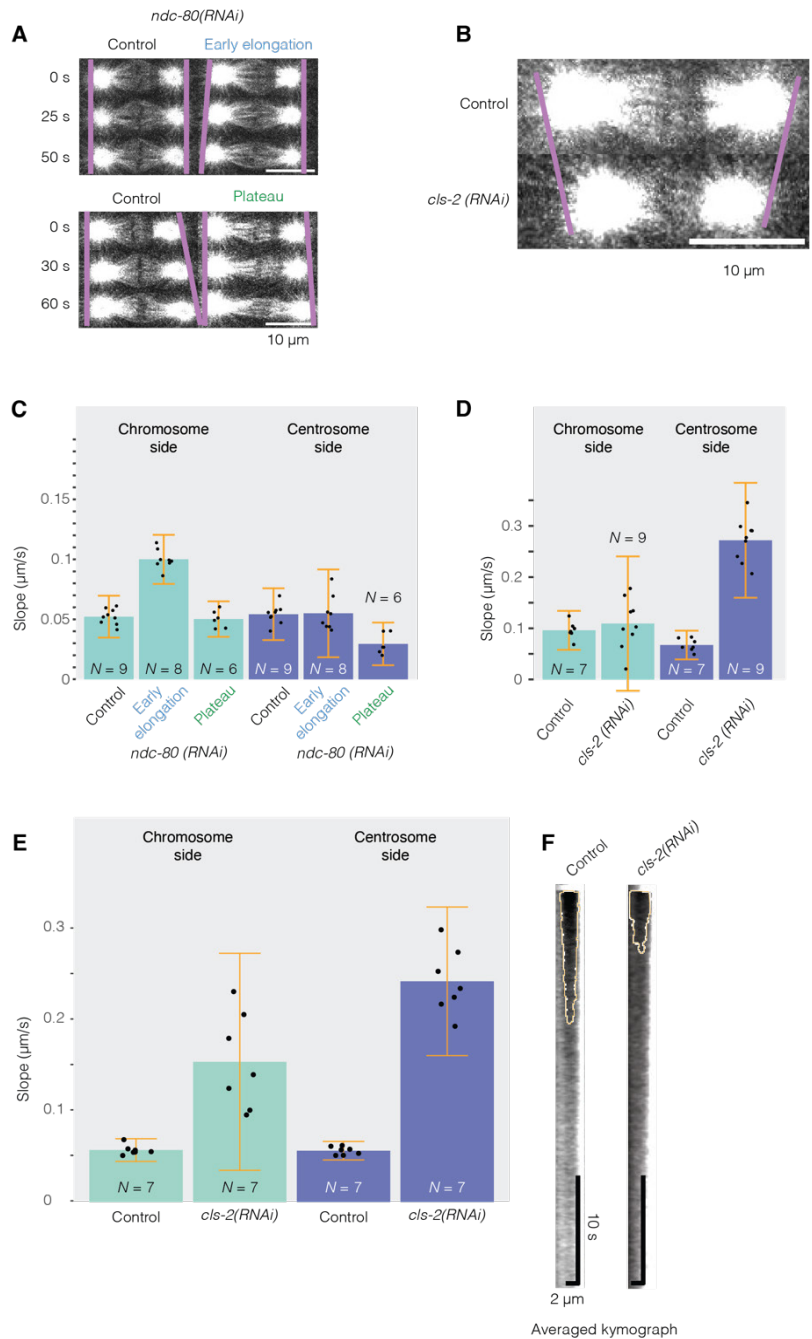

**Figure S7: Microtubule attachment and dynamics at the kinetochore impacted the front velocity. (A, B)** Exemplar micrographs of single GFP::TBB-2$^{\beta\text{-tubulin}}$ embryos used for FRAP experiment and submitted to (A) *ndc-80(RNAi)*, (B) *cls-2(RNAi)* or corresponding control treatments. Times are given from the beginning of the early elongation or plateau phases. **(C, D)** Front velocities after segmenting the bleached region of: (C) $N = 8$ *ndc-80(RNAi)* embryos bleached during precocious spindle elongation, $N = 6$ *ndc-80(RNAi)* bleached during spindle length plateau and their $N = 9$ controls (replica of Fig 2D); (D) $N = 9$ *cls-2(RNAi)* treated embryos and their $N = 7$ control ones. Microtubules were labelled using GFP::TBB-2$^{\beta\text{-tubulin}}$. Black dots represent averages of $N$-1 embryos, leaving out, in turn, each embryo (Methods § Statistics on kymograph front slopes). Bars correspond to means; error bars are estimated using Jackknife resampling. Light blue bars are values on the chromosome side and dark blue on the centrosome side. **(E)** Replica of the experiment (D) using $N = 7$ *cls-2(RNAi)* treated embryos and $N = 7$ control ones. **(F)** Averaged kymographs over the posterior half-spindle for the data presented in (D), with centrosome on the right-hand side. The orange line delineates the bleached region as obtained by our analysis (Methods § Image processing).

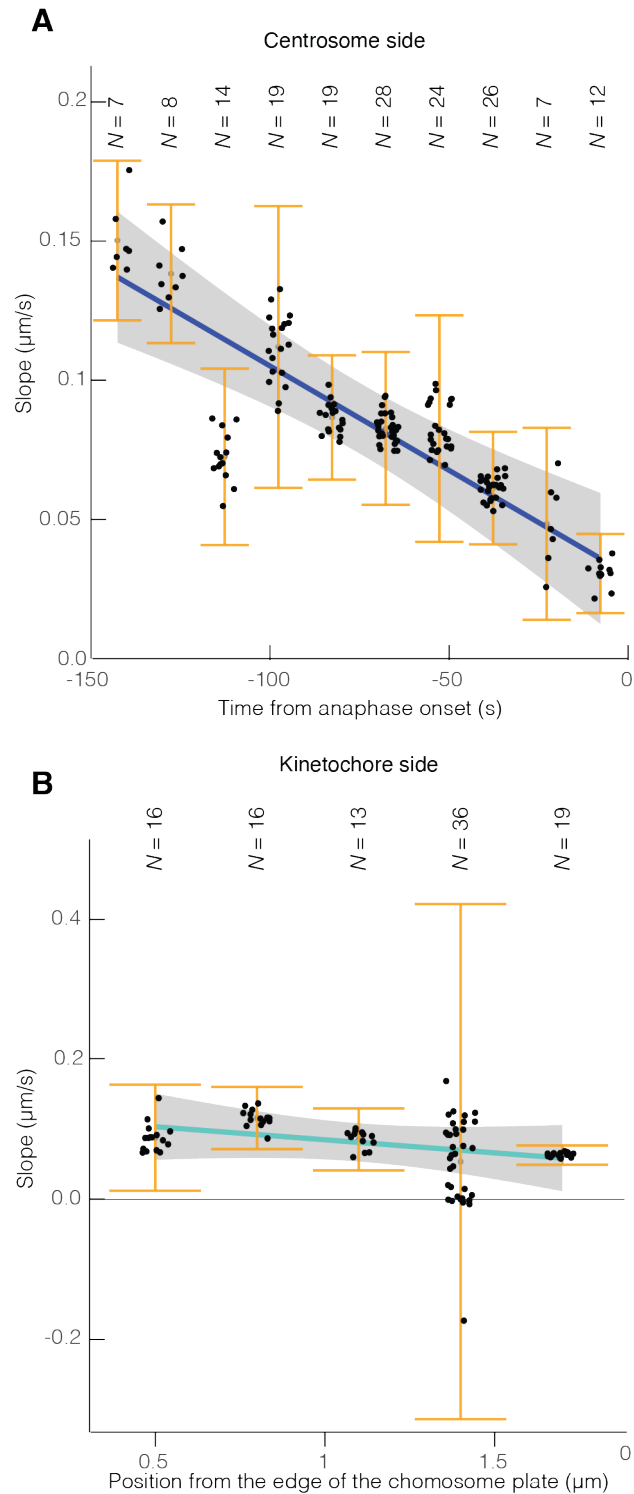

**Figure S8: Slope dependence on timing of bleaching and position along the spindle.** (**A**) Correlation between the bleaching time and the front velocities on the chromosome side. Pearson coefficient read $R = -0.90$ ($p = 0.00037$). (**B**) Correlation between the position of the chromosome-side edge of the initial bleached area and the front velocities measured on the chromosome side. Pearson coefficient read $R = -0.73$ ($p = 0.16$). In both panels, we used non-treated and control RNAi embryos of experiments reported in other figures and segmented the bleached regions. Microtubules were labelled using GFP::TBB-2$^{\beta\text{-tubulin}}$. Black dots represent averages of $N-1$ embryos, leaving out, in turn, each embryo (Methods § Statistics on kymograph front slopes). Bars correspond to means; error bars are estimated using Jackknife resampling. The grey-shaded region corresponds to the confidence interval at 0.95 on the correlation-line coefficients.

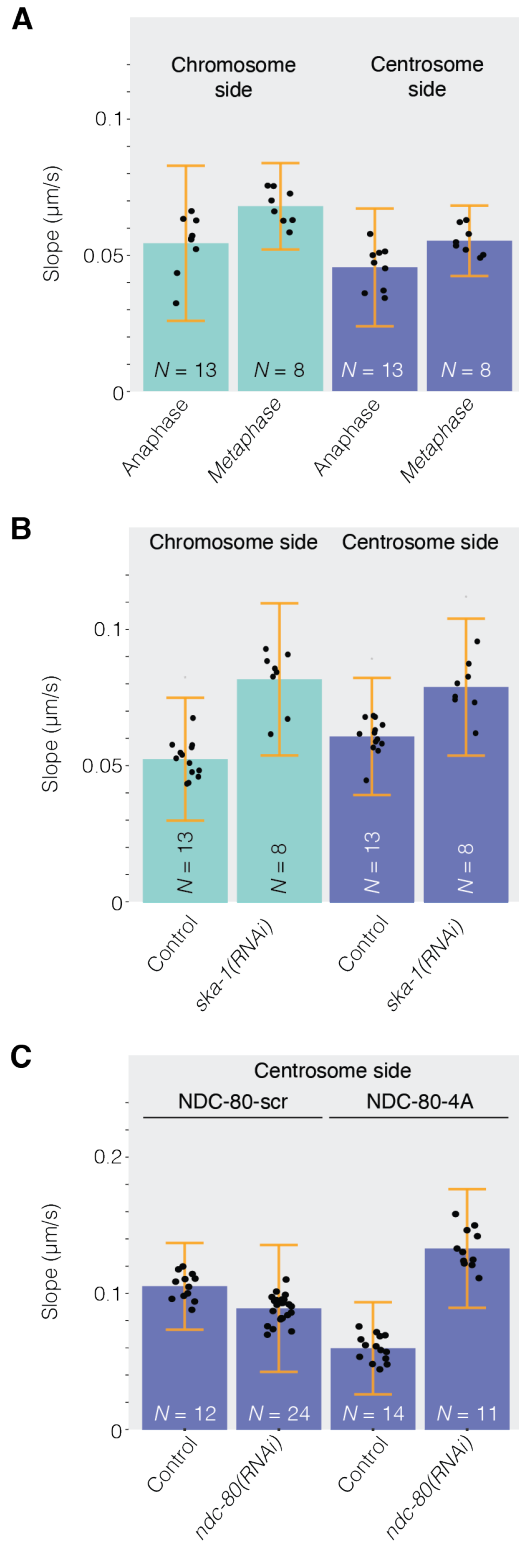

**Figure S9:** (**A**) Front velocities by segmenting the bleached region of $N = 8$ embryos bleached during metaphase and $N = 13$ during anaphase. In all panels, microtubules were labelled using GFP::TBB-2$^{\beta\text{-tubulin}}$. (**B**) Front velocities by segmenting the bleached region of $N = 8$ *ska-1(RNAi)* embryos bleached during late metaphase (-60 to -30 s from anaphase onset) and their $N = 13$ controls. The experiment is a replica of Fig 3B. (**C**) Front velocities on centrosome side by segmenting the bleached region of NDC-80-scr embryos either treated by ($N = 12$) *ndc-80(RNAi)* or ($N = 24$) control; NDC-80-4A embryos either treated by ($N = 14$) *ndc-80(RNAi)* or ($N = 11$) control. Corresponding velocities on the chromosome side are in Fig 3C. Black dots represent averages of $N$-1 embryos, leaving out, in turn, each embryo (Methods § Statistics on kymograph front slopes). Bars correspond to means; error bars are estimated standard errors using Jackknife resampling. Light blue bars are values on the chromosome side and dark blue on the centrosome side.

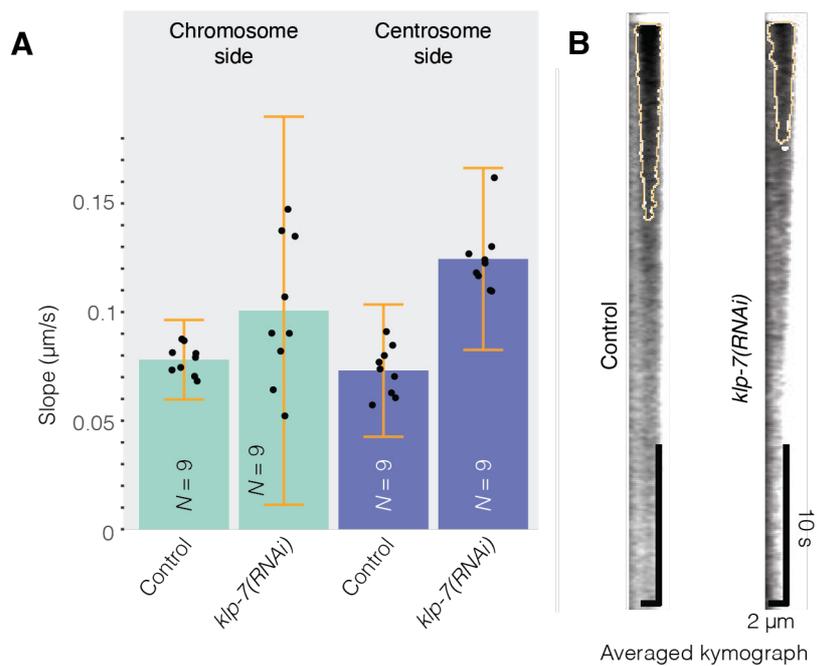

**Figure S10: KLP-7^MCAK was dispensable for the front velocity on the chromosome side.** (A) Front velocities after segmenting the bleached region of $N = 9$ *klp-7(RNAi)* embryos and their $N = 9$ controls. Black dots represent averages of *N*-1 embryos, leaving out, in turn, each embryo (Methods § Statistics on kymograph front slopes). Bars correspond to means; error bars are estimated standard errors using Jackknife resampling. Light blue bars are values on the chromosome side and dark blue on the centrosome side. (B) Averaged kymograph over the posterior half-spindle for the data presented in (A), with centrosome on the right-hand side. The orange line delineates the bleached region as obtained by our analysis (Methods § Image processing). Light blue bars are velocity values measured on the chromosome side and dark blue on the centrosome side.

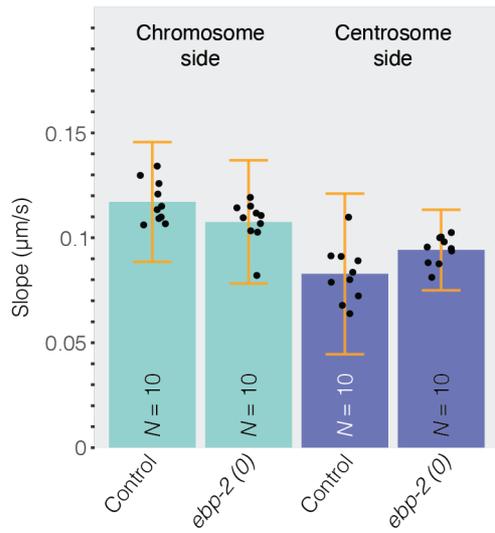

**Figure S11: EBP-2^EB1 was dispensable for the front velocity on the chromosome side.** Front velocities by segmenting the bleached region of $N = 10$ deletion mutant *ebp-2(gk756)* embryos compared to non-treated embryos. Microtubules were labelled using GFP::TBB-2$^{β\text{-tubulin}}$. Black dots represent averages of *N*-1 embryos, leaving out, in turn, each embryo (Methods § Statistics on kymograph front slopes). Bars correspond to means; error bars are estimated using Jackknife resampling. Light blue bars are values on the chromosome side and dark blue on the centrosome side.

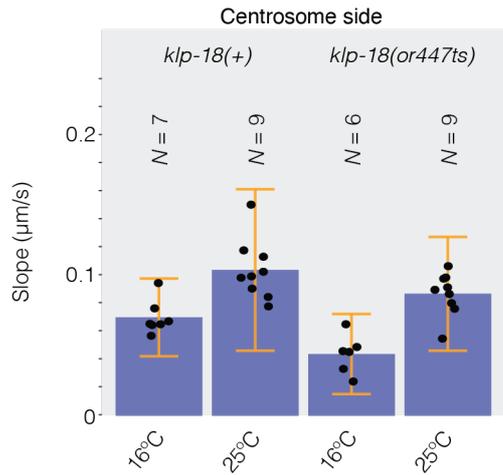

**Figure S12: klp-18$^{Kif15}$ mutation did not impact the centrosome-side front velocity.** Front velocities on the centrosome side by segmenting the bleached region of *klp-18(or447ts)* embryos at ($N = 6$) permissive temperature 16ºC and ($N = 9$) restrictive temperature 25ºC; Corresponding control embryos *klp-18(+)* at ($N = 7$) permissive temperature 16ºC and ($N = 9$) restrictive temperature 25ºC. Microtubules were labelled using GFP::TBB-2$^{β\text{-tubulin}}$. Black dots represent averages of *N*-1 embryos, leaving out, in turn, each embryo (Methods § Statistics on kymograph front slopes). Bars correspond to means; error bars are estimated using Jackknife resampling. Corresponding velocities on the chromosome side are in Fig 4B.

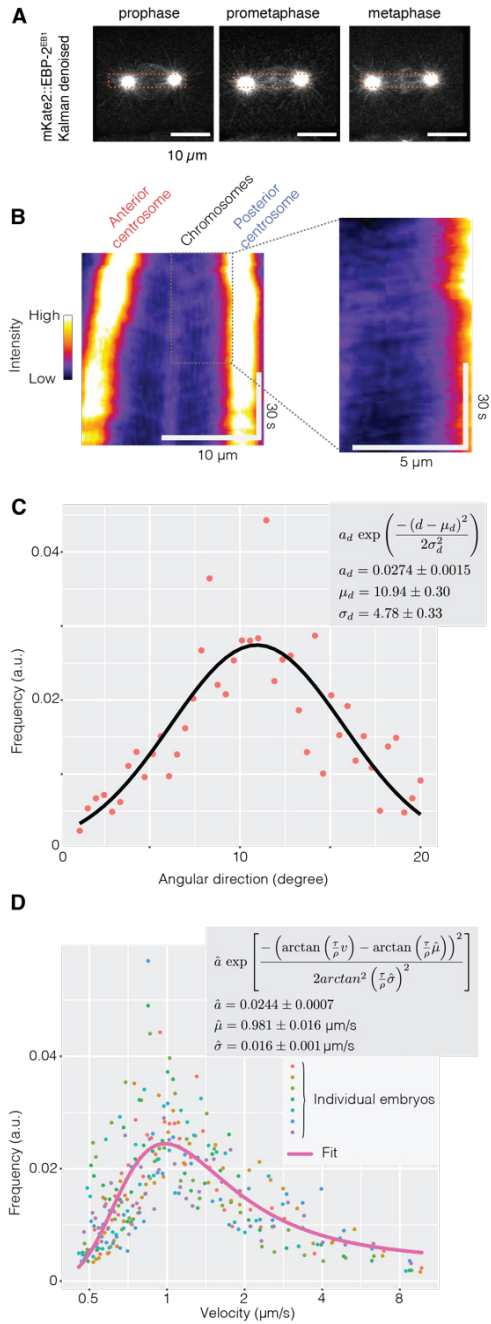

**Figure S13: Measuring microtubule growth rate in the mitotic spindle.** We imaged the spindle of $N = 7$ EBP-2[EB1]::mKate2; GFP::TBB-2[β-tubulin] labelled embryos at 2 Hz and registered the images using the β-tubulin channel. We applied a Kalman denoising. Scale bars indicate 10 μm. (**A**) Typical stills of the EBP-2 (MT plus-ends) channel. The red dashed rectangle depicts the region used to compute the kymograph (Methods § Measuring the growth rate of microtubule plus ends). (**B**) (left part) Corresponding kymograph for a typical embryo and (right part) magnified posterior half-spindle highlighting the traces due to microtubule growth or displacement. (**C**) (line) Exemplar fit of (dots) an individual embryo histogram of comet directions. It corresponds to the angle of the comets measured from the horizontal direction in degrees. We used the usual convention, i.e. positive angles correspond to rotating counter-clockwise. We used a Gaussian function model, with $a$ the fit-estimated normalisation factor, $d$ the direction, $\mu_d$ the estimated direction and $\sigma_d$ the standard deviation. Values obtained by fit with standard errors are reported in the inset. (**D**) (pink line) Global fit of the comet velocity distribution of (coloured dots) individual embryos. We modelled the angular distribution of the comets with a Gaussian and transformed direction into velocity. The equation is reported with $\hat{a}$ the fit-estimated normalisation factor, $v$ the velocity, $\tau$ the cycle time at imaging (different for each embryo), $\rho$ the pixel size (resolution), $\hat{\mu}$ the estimated velocity and $\hat{\sigma}$ the standard deviation. Values obtained by fit with standard errors are reported in the inset.
Page 19 sur 22

## Supplemental Movie

**Movie S1: Tubulin fluorescence recovery after photobleaching in the metaphasic spindle of *C. elegans*.** $N = 10$ GFP::TBB-2$^{\beta\text{-tubulin}}$ labelled embryos were registered on the centrosome-side edge of the photobleached area and combined by computing pixel-wise median (Methods § Image processing, Fig S2). The movie plays in real-time.

**Movie S2: Localisation of KLP-18** in the spindle of a typical *C. elegans* embryo labelled with (left) KLP-18::GFP (middle) EBP-2::mKate2 and (right) their superimposition, acquired on the spinning disk microscope (Methods § Microscopy). The movie plays in real-time.

## Supplemental references